\documentclass[utf8]{FrontiersinHarvard} 

\usepackage{url,hyperref,lineno,microtype,subcaption}
\usepackage[onehalfspacing]{setspace}
\usepackage{amsthm}
\usepackage{amssymb}
\usepackage{xcolor,graphicx,epstopdf,caption,subcaption}
\usepackage{booktabs}
\usepackage{amsmath}

\usepackage{multirow} 
\newtheorem{lemma}{Lemma}
\newtheorem{assumption}{Assumption}

\newtheorem{problem}{Problem}

\def\keyFont{\fontsize{8}{11}\helveticabold }
\def\firstAuthorLast{Jandaghi {et~al.}} 
\def\Authors{Emadodin Jandaghi\,$^{1}$, Mingxi Zhou\,$^{2}$, Paolo Stegagno\,$^{3}$,  Chengzhi Yuan\,$^{1,*}$}
\def\Address{$^{1}$Department of Mechanical, Industrial and Systems Engineering, University of Rhode Island, Kingston, RI, USA

$^{2}$Graduate School of Oceanography, University of Rhode Island, Kingston, RI, USA

$^{3}$Department of Electrical, Computer, and Biomedical Engineering, University of Rhode Island, Kingston, RI, USA}
\def\corrAuthor{Chengzhi Yuan}
\def\corrEmail{cyuan@uri.edu}

\begin{document}
\onecolumn
\firstpage{1}

\title[Running Title]{Adaptive Formation Learning Control for Cooperative AUVs under Complete Uncertainty} 

\author[\firstAuthorLast ]{\Authors} 
\address{} 
\correspondance{} 


\extraAuth{}

\maketitle

\begin{abstract}

\section{}
This paper introduces an innovative two-layer control framework for Autonomous Underwater Vehicles (AUVs), tailored to master completely uncertain nonlinear system dynamics in the presence of challenging hydrodynamic forces and torques. Distinguishing itself from prior studies, which only regarded certain matrix coefficients as unknown while assuming the mass matrix to be known, this research significantly progresses by treating all system dynamics, including the mass matrix, as unknown. This advancement renders the controller fully independent of the robot’s configuration and the diverse environmental conditions encountered. The proposed framework is universally applicable across varied environmental conditions that impact AUVs, featuring a first-layer cooperative estimator and a second-layer decentralized deterministic learning (DDL) controller. This architecture not only supports robust operation under diverse underwater scenarios but also adeptly manages environmental effects such as changes in water viscosity and flow, which influence the AUV's effective mass and damping dynamics. The first-layer cooperative estimator fosters seamless inter-agent communication by sharing crucial system estimates without the reliance on global information, while the second-layer DDL controller utilizes local feedback to precisely adjust each AUV’s trajectory, ensuring accurate formation control and dynamic adaptability. Furthermore, the capability for local learning and knowledge storage through radial basis function neural networks (RBF NNs) is pivotal, allowing AUVs to efficiently reapply learned dynamics following system restarts. Comprehensive simulations validate the effectiveness of our groundbreaking framework, highlighting its role as a major enhancement in the field of distributed adaptive control systems for AUVs. This approach not only bolsters operational flexibility and resilience but is also essential for navigating the unpredictable dynamics of marine environments.
 \keyFont{ \section{Keywords:} Environment-independent Controller, Autonomous Underwater Vehicles (AUV),  Dynamic learning, formation learning control, Multi-agent systems, Neural network control, Adaptive control, Robotics.} 
\end{abstract}

\section{Introduction}
\label{sec:intro}
Robotics and autonomous systems have a wide range of applications, spanning from manufacturing and surgical procedures to exploration in challenging environments \cite{rabiee2024wavelet, ghafoori2024bispectrum, jandaghi2023motion}. However, controlling robots in challenging environments, such as space and underwater, presents significant difficulties due to the unpredictable dynamics in these settings. In the context of underwater exploration, Autonomous Underwater Vehicles (AUVs) have emerged as indispensable tools. These vehicles are not only cost-effective and reliable but also versatile in adapting to the dynamic and often harsh conditions of underwater environments \cite{zhou2023review}. Unlocking the mysteries of marine environments hinges on the effective use of AUVs, making advancements in their control and operation critical.

As the complexity of tasks assigned to AUVs increases, there is a growing need to enhance their operational capabilities. This includes developing sophisticated formation control strategies that allow multiple AUVs to operate in coordination, mimicking collaborative behaviors seen in natural swarms such as fish schools or bird flocks. These strategies are essential for ensuring efficient, stable, and precise operations in varied underwater tasks, ranging from pipeline inspections to seafloor mapping \cite{yan2023formation,hou2009coordinated}. Multi-AUV formation control presents significant challenges due to the intricate nonlinear dynamics and complex interaction behaviors among the AUVs, as well as the uncertain and dynamic nature of underwater environments. Despite these difficulties, the problem has garnered substantial interest from the marine technology and control engineering communities, driven by the extensive applications of AUVs in modern ocean industries, making effective multi-AUV formation control increasingly critical \cite{yan2018coordinated, hou2009coordinated}.

Historically, formation control research has predominantly utilized the behavioral approach \cite{736776, lawton1993nonlinear}, which segmented the overall formation control design problem into multiple subproblems. Each vehicle's control action is then determined by a weighted average of solutions to these subproblems, facilitating decentralized implementation. However, defining the appropriate weighting parameters often presents challenges. Another method, the leader-following approach \cite{cui2010leader, rout2016backstepping}, selected one vehicle as the leader while the rest function as followers. The leader pursues a specified path, and the followers maintain a pre-defined geometric relationship with the leader, enabling control over the formation behavior by designing specific motion behaviors for the leader. Additionally, the virtual structure approach \cite{millan2013formation, ren2008distributed} modified this concept by not assigning a physical leader but instead creating a virtual structure to set the desired formation pattern. This method combines the benefits of the leader-following approach with greater flexibility, as the dynamics of the virtual leader can be tailored for more adaptable formation control designs.

In the field of formation control, several key methodologies have emerged as significant contributors to research. The behavioral approach \cite{736776} addresses the formation control challenge by partitioning the overall design problem into distinct subproblems, where the control solution for each vehicle is computed as a weighted average of these smaller solutions. This technique supports decentralized control but often complicates the calibration of weighting parameters. Conversely, the leader-following strategy \cite{cui2010leader} assigns a primary vehicle as the leader, guiding a formation where subordinate vehicles adhere to a specified geometric arrangement relative to this leader. This approach effectively governs the collective behavior by configuring the leader’s trajectory. Furthermore, the virtual structure method \cite{ren2008distributed} builds upon the leader-following paradigm by substituting the leader with a virtual structure, thereby enhancing the system’s adaptability. This advanced method allows for the flexible specification of virtual leader dynamics, which can be finely tuned to meet diverse operational needs, thus inheriting and enhancing the leader-following model's advantages.

Despite extensive literature in the field, much existing research assumes homogeneous dynamics and certain system parameters for all AUV agents, which is unrealistic in unpredictable underwater environments. Factors such as buoyancy, drag, and varying water viscosity significantly alter system dynamics and behavior. Additionally, AUVs may change shape during tasks like underwater sampling or when equipped with robotic arms, further complicating control. Typically, designing multi-AUV formation control involves planning desired formation paths and developing tracking controllers for each AUV. However, accurately tracking these paths is challenging due to the complex nonlinear dynamics of AUVs, especially when precise models are unavailable. Implementing a fully distributed and decentralized formation control system is also difficult, as centralized control designs become exceedingly complex with larger AUV groups.

In previous work, such as \cite{yuan2017formation}, controllers relied on the assumption of a known mass matrix, which is not practical in real-world applications. Controllers’ dependent on known system parameters can fail due to varying internal forces caused by varying external environmental conditions. The solution lies in developing environment-independent controllers that do not rely on any specific system dynamical parameters, allowing the robot to follow desired trajectories based on desirable signals without any knowledge on system dynamical parameters.

This paper achieves a significant advancement by considering all system dynamics parameters as unknown, enabling a universal application across all AUVs, regardless of their operating environments. This universality is crucial for adapting to environmental variations such as water flow, which increases the AUV’s effective mass via the added mass phenomenon and affects the vehicle's inertia. Additionally, buoyancy forces that vary with depth, along with hydrodynamic forces and torques—stemming from water flow variations, the AUV’s unique shape, its appendages, and drag forces due to water viscosity—significantly impact the damping matrix in the AUV's dynamics. By addressing these challenges, the proposed framework ensures robust and reliable operation across a spectrum of challenging scenarios.

The framework's control architecture is ingeniously divided into a first-layer Cooperative Estimator Observer and a lower-layer Decentralized Deterministic Learning (DDL) Controller. The first-layer observer is pivotal in enhancing inter-agent communication by sharing crucial system estimates, operating independently of any global information. Concurrently, the second-layer DDL controller utilizes local feedback to finely adjust each AUV’s trajectory, ensuring resilient operation under dynamic conditions heavily influenced by hydrodynamic forces and torques by considering system uncertainty completely unknown. This dual-layer setup not only facilitates acute adaptation to uncertain AUV dynamics but also leverages radial basis function neural networks (RBF NNs) for precise local learning and effective knowledge storage. Such capabilities enable AUVs to efficiently reapply previously learned dynamics following any system restarts. This framework not only drastically improves operational efficiency but also significantly advances the field of autonomous underwater vehicle control by laying a robust foundation for future enhancements in distributed adaptive control systems and fostering enhanced collaborative intelligence among multi-agent networks in marine environments. Extensive simulations have underscored the effectiveness of our pioneering framework, demonstrating its substantial potential to elevate the adaptability and resilience of AUV systems under the most demanding conditions.

The rest of the paper is organized as follows: Section~\ref{sec:Preliminaries_Problem_Statement} provides an initial overview of graph theory, Radial Basis Function (RBF) Neural Networks, and the problem statement. The design of the distributed cooperative estimator and the decentralized deterministic learning controller are discussed in Section~\ref{sec:Adaptive}. The formation adaptive control and formation control using pre-learned dynamics are explored in Section~\ref{sec:closedloopformation} and Section~\ref{sec:formationprelearnd}, respectively. Simulation studies are presented in Section~\ref{sec:simulation}, and Section~\ref{sec:conclusion} concludes the paper.

\section{PRELIMINARIES AND PROBLEM STATEMENT}\label{sec:prelim}
\label{sec:Preliminaries_Problem_Statement}
\subsection{Notation and Graph Theory}
Denoting the set of real numbers as \(\mathbb{R}\), we define \(\mathbb{R}^{m \times n}\) as the set of \(m \times n\) real matrices, and \(\mathbb{R}^n\) as the set of \(n \times 1\) real vectors.
The identity matrix is symbolized as \(\mathbf{I}\). 
The vector with all elements being 1 in an \(n\)-dimensional space is represented as \(\mathbf{1}_n\). The sets \(\mathbb{S}_+^n\) and \(\mathbb{S}_-^{n+}\) stand for real symmetric \(n \times n\) and positive definite matrices, respectively. A block diagonal matrix with matrices \(X_1, X_2, \dots, X_p\) on its main diagonal is denoted by \(\text{diag}\{X_1, X_2, \ldots, X_p\}\). 
\(A \otimes B\) signifies the Kronecker product of matrices \(A\) and \(B\). 
For a matrix \(A\), \(\vec{A}\) is the vectorization of \(A\) by stacking its columns on top of each other. For a series of column vectors \(x_1, \dots, x_n\), \(\text{col}\{x_1, \dots, x_n\}\) represents a column vector formed by stacking them together.
Given two integers \(k_1\) and \(k_2\) with \(k_1 < k_2\), \(\mathbf{I}[k_1, k_2] = \{k_1, k_1 + 1, \dots, k_2\}\). For a vector \(x \in \mathbb{R}^n\), its norm is defined as \(|x| := (x^T x)^{1/2}\). For a square matrix \(A\), \(\lambda_i(A)\) denotes its \(i\)-th eigenvalue, while \(\lambda_{\text{min}}(A)\) and \(\lambda_{\text{max}}(A)\) represent its minimum and maximum eigenvalues, respectively.

A directed graph \(G = (V, E)\) comprises nodes in the set \(V = \{1, 2, \dots, N\}\) and edges in \(E \subseteq V \times V\). An edge from node \(i\) to node \(j\) is represented as \((i, j)\), with \(i\) as the parent node and \(j\) as the child node. Node \(i\) is also termed a neighbor of node \(j\). \(N_i\) is considered as the subset of \(V\) consisting of the neighbors of node \(i\). A sequence of edges in \(G\), \((i_1, i_2), (i_2, i_3), \dots, (i_k, i_{k+1})\), is called a path from node \(i_1\) to node \(i_{k+1}\). Node \(i_{k+1}\) is reachable from node \(i_1\). A directed tree is a graph where each node, except for a root node, has exactly one parent. The root node is reachable from all other nodes. A directed graph \(G\) contains a directed spanning tree if at least one node can reach all other nodes. The weighted adjacency matrix of \(G\) is a non-negative matrix \(A = [a_{ij}] \in \mathbb{R}^{N \times N}\), where \(a_{ii} = 0\) and \(a_{ij} > 0 \implies (j, i) \in E\). The Laplacian of \(G\) is denoted as \(L = [l_{ij}] \in \mathbb{R}^{N \times N}\), where \(l_{ii} = \sum_{j=1}^{N} a_{ij}\) and \(l_{ij} = -a_{ij}\) if \(i \neq j\). 
It is established that \(L\) has at least one eigenvalue at the origin, and all nonzero eigenvalues of \(L\) have positive real parts. 
From \cite{ren2005consensus}, \(L\) has one zero eigenvalue and remaining eigenvalues with positive real parts if and only if \(G\) has a directed spanning tree.
\subsection{Radial Basis Function NNs} \label{1b}
The RBF Neural Networks can be described as \(f_{nn}(Z)=\sum_{i = 1}^{N}w_i s_i(Z)=W^{T}S(Z)\), where \(Z\in\Omega_Z\subseteq\mathbb{R}^q\) and \(W=w_1,...,w_N^T\in\mathbb{R}^N\) as input and weight vectors respectively \cite{park1991universal}. \(N\) indicates the number of NN nodes, \(S(Z)=[s_1(||Z-\mu_i||),...,s_N(||Z - \mu_i||)]^T\) with \(s_i(\cdot)\)  is a RBF, and \(\mu_i(i = 1,...,N)\) is distinct points in the state space.
The Gaussian function \(s_i (||Z - \mu_i||)=\operatorname{exp}\left[-\frac{(Z-\mu_i)^T(Z-\mu_i)}{\eta_i^2}\right]\) is generally used for RBF, where \(\mu_i=[\mu_{i1},\mu_{i2},...,\mu_{iN}]^T\) is the center and \(\eta_i\) is the width of the receptive field. The Gaussian function categorized by localized radial basis function \(s\) in the sense that \(s_i (||Z - \mu_i||)\rightarrow 0\) as \(||Z||\rightarrow \infty\).
Moreover, for any bounded trajectory \(Z(t)\) within the compact set \(\Omega_Z\), \(f(Z)\) can be approximated using a limited number of neurons located in a local region along the trajectory\( f(Z) = W^*_{\zeta} S_{\zeta}(Z) + \epsilon_{\zeta}\), where \(\epsilon_{\zeta}\) is the approximation error, with \(\epsilon_{\zeta} = O(\epsilon) = O(\epsilon^*)\), \(S_{\zeta}(Z) = [s_{j_1}(Z), \dots, s_{j_{\zeta}}(Z)]^T \in \mathbb{R}^{N_{\zeta}}\), \(W^*_{\zeta} = [w^*_{j_1}, \dots, w^*_{j_{\zeta}}]^T \in \mathbb{R}^{N_{\zeta}}\), \(N_{\zeta} < N_n\), and the integers \(j_i = j_1, \dots, j_{\zeta}\) are defined by \(|s_{j_i}(Z_p)| > \iota\) (\(\iota > 0\) is a small positive constant) for some \(Z_p \in Z(t)\). This holds if \(\|Z(t) - \xi_{j_i}\| < \epsilon\) for \(t > 0\). The following lemma regarding the persistent excitation (PE) condition for RBF NNs is recalled from \cite{wang2018deterministic}.
\begin{lemma}
    \label{lemma1}
    Consider any continuous recurrent trajectory\footnote{A recurrent trajectory represents a large set of periodic and periodic-like trajectories generated from linear/nonlinear dynamical systems. A detailed characterization of recurrent trajectories can be found in \cite{wang2018deterministic}.} \( Z(t) : [0, \infty) \rightarrow \mathbb{R}^q \). \( Z(t) \) remains in a bounded compact set \( \Omega_Z \subset \mathbb{R}^q \). Then for an RBF Neural Network (NN) \( W^T S(Z) \) with centers placed on a regular lattice (large enough to cover the compact set \( \Omega_Z \)), the regressor subvector \( S_{\zeta}(Z) \) consisting of RBFs with centers located in a small neighborhood of \( Z(t) \) is persistently exciting.
\end{lemma}
\subsection{Problem Statement}
A multi-agent system comprising \(N\) AUVs with heterogeneous nonlinear uncertain dynamics is considered. The dynamics of each AUV can be expressed as \cite{fossen1999guidance}:
\begin{equation}
\begin{aligned}
&\dot{\eta}_i = J_i(\eta_i)\nu_i, \\
&M_i\dot{\nu}_i + C_i(\nu_i)\nu_i + D_i(\nu_i)\nu_i + g_i(\eta_i) + \Delta_i(\chi_i) = \tau_i.
\end{aligned}
\label{eq:auvdynamics}
\end{equation}
In this study, the subscript \(i \in I[1, N]\) identifies each AUV within the multi-agent system. For every \(i \in I[1, N]\), the vector \(\eta_i = [x_i, y_i, \psi_i]^T \in \mathbb{R}^3\) represents the \(i\)-th AUV's position coordinates and heading in the global coordinate frame, while \(\nu_i = [u_i, v_i, r_i]^T \in \mathbb{R}^3\) denotes its linear velocities and angular rate of heading relative to a body-fixed frame. The positive definite inertial matrix \(M_i = M_i^T \in \mathbb{S}_3^+\), Coriolis and centripetal matrix \(C_i(\nu_i)\in \mathbb{R}^{3 \times 3}\), and damping matrix \(D_i(\nu_i)\in \mathbb{R}^{3 \times 3}\) characterize the AUV's dynamic response to motion. The vector \(g_i(\eta_i)\in \mathbb{R}^{3 \times 1}\) accounts for the restoring forces and moments due to gravity and buoyancy. The term \(\Delta_i(\chi_i) \in \mathbb{R}^{3 \times 1}\), with \(\chi_i := \text{col}\{\eta_i, \nu_i\}\), describes the vector of generalized deterministic unmodeled uncertain dynamics for each AUV.

The vector \(\tau_i \in \mathbb{R}^3\) represents the control inputs for each AUV. The associated rotation matrix \(J_i(\eta_i)\) is given by:
\(
J_i(\eta_i) = \begin{bmatrix}
\cos(\psi_i) & \sin(\psi_i) & 0 \\
-\sin(\psi_i) & \cos(\psi_i) & 0 \\
0 & 0 & 1
\end{bmatrix},
\)
Unlike previous work \cite{yuan2017formation} which assumed known values for the AUV's inertia matrix and rotation matrix, this study considers not only the matrix coefficients \(C_i(\nu_i)\), \(D_i(\nu_i)\), \(g_i(\eta_i)\), and \(\Delta_i(\chi_i)\) but also treats the inertia matrix and rotation matrix as completely unknown and uncertain. When all system parameters are considered unknown, external forces do not influence the controller, making it universally applicable to any AUV regardless of shape, weight, and environmental conditions, including variations in water flow or depth that typically affect damping and inertia forces during operation.

In the context of leader-following formation tracking control, the following virtual leader dynamics generates the tracking reference signals:
\begin{equation}
    \begin{aligned}
        \dot{\chi}_0 = A_0\chi_0
        \label{eq:leaderdynamics}
    \end{aligned}
\end{equation}

With ``0'' marking the leader node, the leader state \(\chi_0 := \operatorname{col}\{\eta_0, \nu_0\}\) with \(\eta_0 \in \mathbb{R}^3\) and \(\nu_0 \in \mathbb{R}^3\), \(A_0 \in \mathbb{R}^{3 \times 3}\) is a constant matrix available only to the leader’s neighboring AUV agents.

Considering the system dynamics of multiple AUVs \eqref{eq:auvdynamics} along with the leader dynamics \eqref{eq:leaderdynamics}, we establish a non-negative matrix \(A = [a_{ij}]\), where \(i, j \in I[0, N]\) such that for each \(i \in I[1, N]\), \(a_{i0} > 0\) if and only if agent \(i\) has access to the reference signals \(\eta_0\) and \(\nu_0\). All remaining elements of \(A\) are arbitrary non-negative values, such that \(a_{ii} = 0\) for all \(i\). Correspondingly, we establish \(G = (V, E)\) as a directed graph derived from \(A\), where \(V = \{0, 1, \dots, N\}\) designates node \(0\) as the leader, and the remaining nodes correspond to the \(N\) AUV agents. We proceed under the following assumptions:

\begin{assumption}\label{assump1}
All the eigenvalues of \(A_0\) in the leader’s dynamics \eqref{eq:leaderdynamics} are located on the imaginary axis.
\end{assumption}
\begin{assumption}\label{assump2}
The directed graph \(G\) contains a directed spanning tree with the node \(0\) as its root.
\end{assumption}
Assumption~\ref{assump1} is crucial for ensuring that the leader dynamics produce stable, meaningful reference trajectories for formation control. It ensures that all states of the leader, represented by \(\chi_0 = \text{col}\{\eta_0, \nu_0\}\), remain within the bounds of a compact set \(\Omega_0 \subset \mathbb{R}^6\) for all \(t \geq 0\). The trajectory of the system, starting from \(\chi_0(0)\) and denoted by \(\phi_0(\chi_0(0))\), generates periodic signal. This periodicity is essential for maintaining the Persistent Excitation (PE) condition, which is pivotal for achieving parameter convergence in Distributed Adaptive (DA) control systems. Modifications to the eigenvalue constraints on \(A_0\) mentioned in Assumption~\ref{assump1} may be considered when focusing primarily on formation tracking control performance, as discussed later.

\begin{figure}[h!]
\begin{center}
\includegraphics[width=10cm]{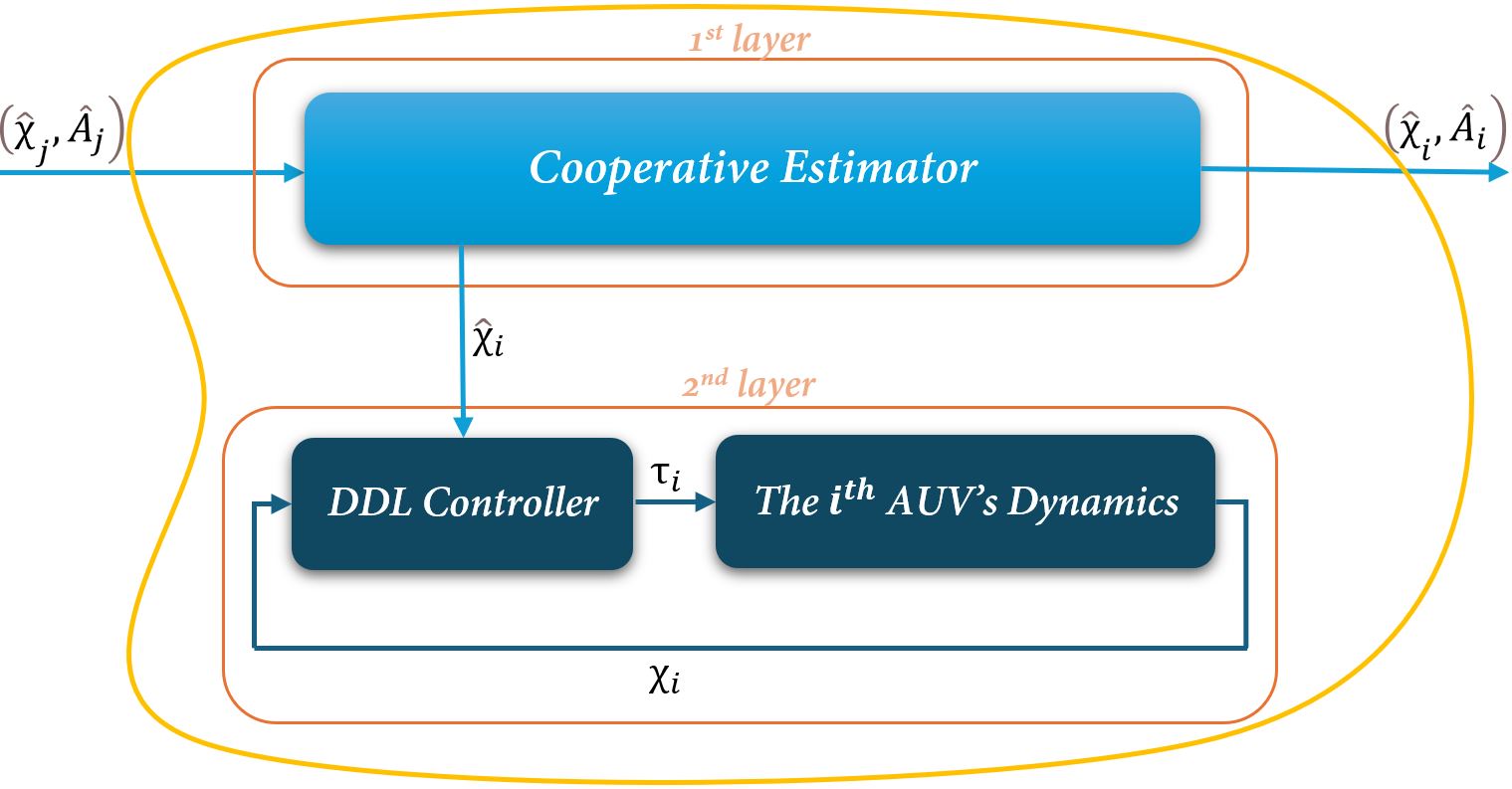}
\end{center}
\caption{Proposed two-layer distributed controller architecture for each AUVs}
\label{fig:Twolayer}
\end{figure}

Additionally, Assumption~\ref{assump2} reveals key insights into the structure of the Laplacian matrix \(L\) of the network graph \(G\). Let \(\Psi\) be an \(N \times N\) non-negative diagonal matrix where each \(i\)-th diagonal element is \(a_{i0}\) for \(i \in I[1, N]\). The Laplacian \(L\) is formulated as:
\begin{equation}
    L = \begin{bmatrix} 
    \sum_{j=1}^{N} a_{0j} & -[a_{01}, \ldots, a_{0N}] \\
    -\Psi \mathbf{1}_N & H
    \end{bmatrix},
    \label{laplacian}
\end{equation}
where \(a_{0j} > 0\) if \((j, 0) \in E\) and \(a_{0j} = 0\) otherwise. This results in \(\mathbf{H}\mathbf{1}_N = \Psi \mathbf{1}_N\) since \(L\mathbf{1}_{N+1} = 0\). As cited in \cite{su2011cooperative}, all nonzero eigenvalues of \(\mathbf{H}\), if present, exhibit positive real parts, confirming \(\mathbf{H}\) as nonsingular under Assumption~\ref{assump2}.
\begin{problem}
    \label{problem1}
    In the context of a multi-AUV system \eqref{eq:auvdynamics} integrated with virtual leader dynamics \eqref{eq:leaderdynamics} and operating within a directed network topology \(G\), the aim is to develop a distributed NN learning control protocol that leverages only local information. The specific goals are twofold:
    
1) \textbf{Formation Control:} Each of the \(N\) AUV agents will adhere to a predetermined formation pattern relative to the leader, maintaining a specified distance from the leader's position \(\eta_0\).

2) \textbf{Decentralized Learning:} The nonlinear uncertain dynamics of each AUV will be identified and learned autonomously during the formation control process. The insights gained from this learning process will be utilized to enhance the stability and performance of the formation control system.
\end{problem}
\newtheorem{remark}{Remark}
\begin{remark} 
    \label{remark1}
The leader dynamics described in Equation \eqref{eq:leaderdynamics} are designed as a neutrally stable LTI system. This design choice facilitates the generation of sinusoidal reference trajectories at various frequencies which is essential for effective formation tracking control. This approach to leader dynamics is prevalent in the literature on multiagent leader-following distributed control systems like \cite{yuan2017leader} and \cite{10637197}.
\end{remark}
\begin{remark} 
    \label{remark2}
    It is important to emphasize that the formulation assumes formation control is required only within the horizontal plane, suitable for AUVs operating at a constant depth, and that the vertical dynamics of the 6-DOF AUV system, as detailed in \cite{prestero2001development}, are entirely decoupled from the horizontal dynamics.
\end{remark}
As shown in Fig~\ref{fig:Twolayer}, a two-layer hierarchical design approach is proposed to address the aforementioned challenges. The first layer,the Cooperative Estimator, enables information exchange among neighboring agents. The second layer, known as the Decentralized Deterministic Learning (DDL) controller, processes only local data from each individual AUV. The subsequent two sections will discuss the development of the DA observer within the first layer and the formulation of the DDL control strategy, respectively.
\section{two-layer distributed controller architecture}
\label{sec:Adaptive}
\subsection{First Layer: Cooperative Estimator}
In the context of leader-following formation control, not all AUV agents may have direct access to the leader’s information, including tracking reference signals (\(\chi_0\)) and the system matrix (\(A_0\)). This necessitates collaborative interactions among the AUV agents to estimate the leader’s information effectively. Drawing on principles from multiagent consensus and graph theories \cite{ren2008distributed}, we propose to develop a distributed adaptive observer for the AUV systems as:
\begin{equation}
\dot{\hat\chi}_i(t) = A_i(t)\hat\chi_i(t) + \beta_{i1} \sum_{j=0}^{N} a_{ij} (\hat\chi_j(t) - \hat\chi_i(t)), \quad \forall i \in I[1, N]
\label{eqn:DAO}
\end{equation}
The observer states for each \(i\)-th AUV, denoted by \(\hat{\chi}_i = [\hat{\eta}_i, \hat{\nu}_i]^T \in \mathbb{R}^6\), aim to estimate the leader's state, \(\chi_0 = [\eta_0, \nu_0]^T \in \mathbb{R}^6\). As \( t \to \infty \), these estimates are expected to converge, such that \(\hat\eta_i\) approaches \(\eta_0\) and \(\hat\nu_i\) approaches \(\nu_0\), representing the leader's position and velocity, respectively. The time-varying parameters \(A_0(t)\) for each observer are dynamically adjusted using a cooperative adaptation law:
\begin{equation}
\dot{\hat{A}}_i(t) = \beta_{i2} \sum_{j=0}^{N} a_{ij} (\hat{A}_j(t) - \hat{A}_i(t)), \quad \forall i \in I[1, N]
\label{eq:adaptationlaw}
\end{equation}
\( \hat{A}_i \) are used to estimate the leader's system matrix \( A_0 \) while both are \(\in \mathbb{R}^{6*6}\). The constants \( \beta_{i1} \) and \( \beta_{i2} \) are all positive numbers and are subject to design.
\begin{remark} 
    \label{remark3}
    Each AUV agent in the group is equipped with an observer configured as specified in equations \eqref{eqn:DAO} and \eqref{eq:adaptationlaw}, comprising two state variables, \(\chi_i\) and \(A_i\). For each \(i \in I[1, N]\), \(\chi_i\) estimates the virtual leader’s state \(\chi_0\), while \(A_i\) estimates the leader’s system matrix \(A_0\). The real-time data necessary for operating the \(i\)-th observer includes: (1) the estimated state \(\hat\chi_j\) and matrix \(\hat{A_j}\), obtained from the \(j\)-th AUV itself, and (2) the estimated states \(\chi_j\) and matrices \(A_j\) for all \(j \in N_i\), obtains from the \(j\)-th AUV’s neighbors. Note that in equations \eqref{eqn:DAO} and \eqref{eq:adaptationlaw}, if \(j \notin N_i\), then \(a_{ij} = 0\), indicating that the \(i\)-th observer does not utilize information from the \(j\)-th AUV agent. This configuration ensures that the proposed distributed observer can be implemented in each local AUV agent using only locally estimated data from the agent itself and its immediate neighbors, without the need for global information such as the size of the AUV group or the network interconnection topology.
\end{remark}
By defining the estimation error for the state and the system matrix for agent \(i\) as \( \tilde{\chi}_i = \hat{\chi}_i - \chi_0 \) and \( \tilde{A}_i = \hat{A}_i - A_0 \), respectively, we derive the error dynamics:
\begin{equation*}
\begin{aligned}
\dot{\tilde{\chi}}_i(t) &= \hat{A}_i(t)\hat{\chi}_i(t) + A_0\chi_0(t) + \beta_{i1} \sum_{j=0}^{N} a_{ij} (\hat{\chi}_j(t) - \hat{\chi}_i(t)) \\ 
 &= \hat{A}_{i}(t) \hat {\chi }_{i}\left ({t}\right) - A_{0}\hat {\chi }_{i}\left ({t}\right) + A_{0}\hat {\chi }_{i}\left ({t}\right ) - A_{0}\chi _{0}\left ({t}\right ) + \beta _{1}\sum _{j=0}^{N}a_{ij} \left ({\hat {\chi }_{j}\left ({t}\right ) - \chi _{0}\left ({t}\right ) + \chi _{0}\left ({t}\right ) - \hat {\chi }_{i}\left ({t}\right )}\right ) \\ &= A_{0}\tilde {\chi }_{i}\left ({t}\right ) + \tilde {A}_{i}\left ({t}\right )\tilde {\chi }_{i}\left ({t}\right ) + \tilde {A}_{i}\left ({t}\right )\chi _{0}\left ({t}\right ) + \beta _{1}\sum _{j=0}^{N}a_{ij} \left ({\tilde {\chi }_{j}\left ({t}\right ) - \tilde {\chi }_{i}\left ({t}\right )}\right )\\ 
\dot{\tilde{A}}_i(t) &= \beta_{i2} \sum_{j=0}^{N} a_{ij} - (\tilde{A}_j(t) - \tilde{A}_i(t)) , \quad \forall i \in I[1, N].
\end{aligned}
\end{equation*}
Define the collective error states and adaptation matrices: \(\tilde{\chi} = \text{col}\{\tilde{\chi}_1, \ldots, \tilde{\chi}_N\}\) for the state errors,
\(\tilde{A} = \text{col}\{\tilde{A}_1, \ldots, \tilde{A}_N\}\) for the adaptive parameter errors,
\(\tilde{A}_b = \text{diag}\{\tilde{A}_1, \ldots, \tilde{A}_N\}\) representing the block diagonal of adaptive parameters,
\(B_{\beta_1} = \text{diag}\{\beta_{11}, \ldots, \beta_{N1}\}\) and
\(B_{\beta_2} = \text{diag}\{\beta_{12}, \ldots, \beta_{N2}\}\) for the diagonal matrices of design constants. 
With these definitions, the network-wide error dynamics can be expressed as:
\begin{equation}
\begin{aligned}
\dot{\tilde{\chi}}(t) &= ((I_N \otimes A_0) - B_{\beta_1}(H \otimes I_{2n})) \tilde{\chi}(t) + (\tilde{A}_b(t) \otimes I_{2n}) \tilde{\chi}(t) + \tilde{A}_b(t) (1_N \otimes \chi_0(t)), \\
\dot{\tilde{A}}(t) &= -B_{\beta_2}(H \otimes I_n) \tilde{A}(t).
\end{aligned}
\label{errorsystem}
\end{equation}
\newtheorem{theorem}{Theorem}
\begin{theorem}
\label{theorem1} Consider the error system equations \eqref{errorsystem}. Under Assumptions~\ref{assump1} and ~\ref{assump2}, and given that \(\beta_1, \beta_2 > 0\), it follows that for all \(i \in I[1, N]\) and for any initial conditions \(\chi_0(0), \chi_i(0), A_i(0)\), the error dynamics of the adaptive parameters and the states will converge to zero exponentially. Specifically, \(\lim_{t \to \infty} \tilde{A}_i(t) = 0\) and \(\lim_{t \to \infty} \tilde{\chi}_i(t) = 0\).
\end{theorem}
This convergence is facilitated by the independent adaptation of each agent's parameters within their respective error dynamics, represented by the block diagonal structure of \(\tilde{A}_b\) and control gains \(B_{\beta_1}\) and \(B_{\beta_2}\). These matrices ensure that each agent's parameter updates are governed by local interactions and error feedback, consistent with the decentralized control framework.

\textbf{Proof:} We begin by examining the estimation error dynamics for \(\tilde{A}\) as presented in equation \ref{errorsystem}. This can be rewritten in the vector form:
\begin{equation}
\begin{aligned}
\dot{\vec{\tilde{A}}}_{0}(t) = -\beta_{2}(I_{6} \otimes (\mathcal{H} \otimes I_{3})) \vec{\tilde{A}}_{0}(t) 
\label{eq:8}
\end{aligned}
\end{equation}
Under assumption \ref{assump2}, all eigenvalues of \( \mathcal{H} \) possess positive real parts according to \cite{su2011cooperative}. Consequently, for any positive \(\beta_2 > 0\), the matrix \(-\beta_2(I_6 \otimes (H \otimes I_3))\) is guaranteed to be Hurwitz, Which implies the exponential stability of system \eqref{eq:8}. Hence, it follows that \(\lim_{t \to \infty} \vec{\tilde{A}}_0(t) = 0\) exponentially, leading to \(\lim_{t \to \infty} \tilde{A}_{i0}(t) = 0\) exponentially for all \(i \in I[1, N]\).

Next, we analyze the error dynamics for \(\chi_0\) in equation \eqref{errorsystem}. Based on the previous discussions, We have \(\lim_{t \to \infty} \tilde{A}_b(t) = 0\) exponentially, and the term \(\tilde{A}_b(t)(1_N \otimes \chi_0(t))\) will similarly decay to zero exponentially. Based on \cite{cai2015cooperative}, if the system defined by
\begin{equation} 
\dot{\tilde{\chi}}_{0}(t) = \left((I_{N} \otimes A_{0}) - \beta_{1}(\mathcal{H} \otimes I_{6})\right) \tilde{\chi}_{0}(t) 
\label{9}
\end{equation}
is exponentially stable, then \(\lim_{t \to \infty} \chi_0(t) = 0\) exponentially. With assumption \ref{assump1}, knowing that all eigenvalues of \(A_0\) have zero real parts, and since \(\mathcal{H}\) as nonsingular with all eigenvalues in the right-half plane, system \eqref{9} is exponentially stable for any positive \(\beta_1 > 0\). Consequently, this ensures that \(\lim_{t \to \infty} \chi_0(t) = 0\), i.e., \(\lim_{t \to \infty} \chi_{i0}(t) = 0\) exponentially for all \(i \in I[1, N]\).

Now, each individual agent can accurately estimate both the state and the system matrix of the leader through cooperative observer estimation \eqref{eqn:DAO} and \eqref{eq:adaptationlaw}. This information will be utilized in the DDL controller design for each agent's second layer, which will be discussed in the following subsection.

\subsection{Second Layer: Decentralized Deterministic Learning Controller}
\label{SecondLayer}
To fulfill the overall formation learning control objectives, in this section, we develop the DDL control law for the multi-AUV system defined in \eqref{eq:auvdynamics}. We use \(d^*_i\) to denote the desired distance between the position of the \(i\)-th AUV agent \(\eta_i\) and the virtual leader's position \(\eta_0\). Then, the formation control problem is framed as a position tracking control task, where each local AUV agent's position \(\eta_i\) is required to track the reference signal \(\eta_{d,i} := \eta_0 + d^*_i\).

Besides, due to the inaccessibility of the leader’s state information \(\chi_0\) for all AUV agents, the tracking reference signal \(\hat\eta_{d,i} := \hat\eta_{0,i} + d^*_i\) is employed instead of the reference signal \(\hat\eta_{d,i}\). As established in theorem \ref{theorem1}, \(\hat\eta_{d,i}\) is autonomously generated by each local agent and will exponentially converge to \(\eta_{d,i}\). This ensures that the DDL controller is feasible and the formation control objectives are achievable for all \(i \in I[1, N]\) using \(\hat\eta_{d,i}\).

To design the DDL control law that addresses the formation tracking control and the precise learning of the AUVs' complete nonlinear uncertain dynamics at the same time, we will integrate renowned backstepping adaptive control design method outlined in \cite{krstic1995nonlinear} along with techniques from \cite{wang2009deterministic} and \cite{yuan2017formation} for deterministic learning using RBF NN. Specifically, for the \(i\)-th AUV agent described in system \eqref{eq:auvdynamics}, we define the position tracking error as \(z_{1,i} = \eta_i - \hat{\eta}_{d,i}\) for all \(i \in I[1, N]\). Considering \( J_i(\eta_i) J_i^T(\eta_i) = I \) for all \(i \in I[1, N]\), we proceed to:
\begin{equation} 
\begin{aligned}
    \dot{z}_{1,i} = J_{i}(\eta_i)\nu_i - \dot {\hat {\eta }}_i, \quad \forall i\in \mathbf {I}[1,N].
    \label{eq;positiontrackingder}
\end{aligned}
\end{equation}
To frame the problem in a more tractable way, we assume \(\nu_i\) as a virtual control input and \(\alpha_i\) as a desired virtual control input in our control strategy design, and by implementing them in the above system we have:
\begin{equation}
\begin{aligned}
z_{2,i} &= \nu_i - \alpha_i, \\
\alpha_i &= J_i^T(\eta_i) (-K_{1,i}z_{1,i} + \dot{\hat{\eta}}_i), \quad \forall i \in \mathbf{I}[1, N].
\label{eq:virtualcontrolinput}
\end{aligned}
\end{equation}
A positive definite gain matrix \( K_{1,i} \in \mathbb{S}_3^+ \) is used for tuning the performance. Substituting \(\nu_i = z_{2,i} + \alpha_i\) into (\eqref{eq;positiontrackingder}) yields:
\begin{equation}
\begin{aligned}
\dot{z}_{1,i} &= J_i(\eta_i)z_{2,i} - K_{1,i}z_{1,i}, \quad \forall i \in \mathbf{I}[1, N].
\end{aligned}
\end{equation}
Now we derive the first derivatives of the virtual control input and the desired control input as follows:
\begin{equation}
\begin{aligned}
\dot{z}_{2,i} &= \dot{\nu}_i - \dot{\alpha}_i \\
&= M_i^{-1}(-C_i(\nu_i)\nu_i - D_i(\nu_i)\nu_i - g_i(\eta_i) - \Delta_i(\chi_i) + \tau_i) - \dot{\alpha}_i, 
\label{eq:vritualcontinput}
\end{aligned}
\end{equation}
\begin{equation}
\begin{aligned}
\dot{\alpha}_i &= \dot{J}_i^T(\eta_i)(-K_{1,i}z_{1,i} + \dot{\hat{\eta}}_i) + J_i^T(\eta_i)(K_{1,i}\dot{\hat{\eta}}_i - K_{1,i}J_i(\eta_i)\nu_i + \ddot{\hat{\eta}}_i). \quad \forall i \in \mathbf{I}[1, N],
\label{eq:desiredvritualcontinput}
\end{aligned}
\end{equation}
As previously discussed, unlike earlier research that only identified the matrix coefficients \(C_i(\nu_i)\), \(D_i(\nu_i)\), \(g_i(\eta_i)\), and \(\Delta_i(\chi_i)\) as unknown system nonlinearities while assuming the mass matrix \(M_i\) to be known, this work advances significantly by also considering \(M_i\) as unknown. Consequently, all system dynamic parameters are treated as completely unknown, making the controller fully independent of the robot's configuration—such as its dimensions, mass, or any appendages—and the uncertain environmental conditions it encounters, like depth, water flow, and viscosity. This independence is critical as it ensures that the controller does not rely on predefined assumptions about the dynamics, aligning with the main goal of this research. To address these challenges, we define a unique nonlinear function \(F_i(Z_i)\) that encapsulates all nonlinear uncertainties as follows:
\begin{equation}
\begin{aligned}
    F_i(Z_i) = M_i\dot{\alpha}_i + C_i(\nu_i)\nu_i + D_i(\nu_i)\nu_i + g_i(\eta_i) + \Delta_i(\chi_i),
    \label{eq:nonlinearfunc}
\end{aligned}
\end{equation}
where \(F_i(Z_i) = [f_{1,i}(Z_i), f_{2,i}(Z_i), f_{3,i}(Z_i)]^T\) and \(Z_i = \text{col}\{\eta_i, \nu_i\} \in \Omega_{Z_i} \subset \mathbb{R}^6\), with \(\Omega_{Z_i}\) being a bounded compact set.
We then employ the following RBF NNs to approximate \(f_{k,i}\) for all \(i \in \mathbf{I}[1, N]\) and \(k \in \mathbf{I}[1, 3]\):
\begin{equation}
\begin{aligned}
    f_{k,i}(Z_i) = W_{k,i}^{*T}S_{ki}(Z_i) + \epsilon _{k,i}(Z_i),
    \label{15}
\end{aligned}
\end{equation}
where \(W_{k,i}^{*}\) is the ideal constant NN weights, and \(\epsilon_{k,i}(Z_i)\) is the approximation error \(\epsilon^*_{k,i} > 0\) for all \(i \in \mathbf{I}[1, N]\) and \(k \in \mathbf{I}[1, 3]\), which satisfies \(|\epsilon_{k,i}(Z_i)| \leq \epsilon^*_{k,i}\). This error can be made arbitrarily small given a sufficient number of neurons in the network.
A self-adaptation law is designed to estimate the unknown \(W_{k,i}^*\) online. We aim to estimate \(W_{k,i}^*\) with \(\hat{W}_{k,i}\) by constructing the DDL feedback control law as follows:
\begin{equation}
    \begin{aligned}
        \tau _{i} = -J_{i}^{T}(\eta _{i})z_{1,i} - K_{2,i}z_{2,i} + \hat {W}_{k,i}^{T}S_{k,i}(Z_{i})
        \label{eq:ddlfeedback}
    \end{aligned}
\end{equation}
To approximate the unknown nonlinear function vector \(F_i(Z_i)\) in \eqref{eq:nonlinearfunc} along the trajectory \(Z_i\) within the compact set \(\Omega_{Z_i}\), we use:
\begin{equation}
\hat{W}_i^T S_i^F(Z_i) = 
\begin{bmatrix}
\hat{W}_{1,i}^T S_{1,i}(Z_i) \\
\hat{W}_{2,i}^T S_{2,i}(Z_i) \\
\hat{W}_{3,i}^T S_{3,i}(Z_i)
\end{bmatrix}.
\end{equation}
Also, \(K_{2,i} \in \mathbb{S}_3^+\) is a feedback gain matrix that can be tuned to achieve the desired performance. From \eqref{eq:auvdynamics} and \eqref{eq:ddlfeedback} we have:
\begin{equation}
    \begin{aligned}
       M_i\dot{\nu}_i + C_i(\nu_i)\nu_i + D_i(\nu_i)\nu_i + g_i(\eta_i) + \Delta_i(\chi_i) = \tau _{i} = -J_{i}^{T}(\eta _{i})z_{1,i} - K_{2,i}z_{2,i} + \hat {W}_{k,i}^{T}S_{k,i}(Z_{i})
        \label{eq:18}
    \end{aligned}
\end{equation}
By subtracting \(W_{k,i}^{*T}S_{k,i}(Z_i) + \epsilon_{k,i}(Z_i)\) from both sides and considering Equations \eqref{eq:virtualcontrolinput} and \eqref{eq:nonlinearfunc}, we define \(\tilde{W}_{k,i} := \hat{W}_{k,i} - W_{k,i}^*\), leading to:
\begin{equation}
    \begin{aligned}
   \dot{z}_{2,i} = M_i^{-1} (-J_{i}^{T}(\eta _{i})z_{1,i} - K_{2,i}z_{2,i} + \hat {W}_{k,i}^{T}S_{k,i}(Z_{i}) - \epsilon _{k,i}(Z_i))
        \label{eq:19}
    \end{aligned}
\end{equation}
For updating \( \hat{W}_{k,i}\) online, a robust self-adaptation law is constructed using the \(\sigma\)-modification technique \cite{ioannou1996robust} as follows:
\begin{equation}
    \begin{aligned}
        \dot {\hat{W}}_{k,i} = -\Gamma_{k,i} ({S_{k,i}(Z_{i})z_{2k,i} + \sigma _{ki}\hat {W}_{k,i}})
        \label{eq:robustlaw}
    \end{aligned}
\end{equation}
where \(z_{2,i} = [z_{21,i}, z_{22,i}, z_{23,i}]^T\), \(\Gamma_{k,i} = \Gamma_{k,i}^T > 0\), and \(\sigma_{k,i} > 0\) are free parameters to be designed for all \(i \in \mathbf{I}[1, N]\) and \(k \in \mathbf{I}[1, 3]\).
Integrating equations \eqref{eq;positiontrackingder}, \eqref{eq:ddlfeedback} and \eqref{eq:robustlaw} yields the following closed-loop system:
\begin{equation}
\left\{
\begin{aligned}
    \dot{z}_{1,i} &= -K_{1,i}z_{1,i} + J_i(\eta_i)z_{2,i}, \\
    \dot{z}_{2,i} &= M_i^{-1}(-J_i^T(\eta_i)z_{1,i} - K_{2,i}z_{2,i} + \hat {W}_{ki}^{T}S_{ki}(Z_{i}) - \epsilon_{k,i}(Z_i)), \\
    \dot{\tilde{W}}_{k,i} &= -\Gamma_{k,i}(S_{k,i}(Z_i)z_{2,k,i} + \sigma_{k,i}\hat{W}_{k,i}),
\end{aligned}
\right.
\label{eq:closedloopdynamics}
\end{equation}
where, for all \(i \in \mathbf{I}[1, N]\) and \(k \in \mathbf{I}[1, 3]\), \(\tilde{W}_i^T S_i(Z_i) = [\tilde{W}_{1,i}^T S_{1,i}(Z_i), \tilde{W}_{2,i}^T S_{2,i}(Z_i), \tilde{W}_{3,i}^T S_{3,i}(Z_i)]^T\), and \(\epsilon_i(Z_i) = [\epsilon_{1,i}(Z_i), \epsilon_{2,i}(Z_i), \epsilon_{3,i}(Z_i)]^T\).
\begin{remark} 
    \label{remark4}
Unlike the first-layer DA observer design, the second-layer control law is fully decentralized for each local agent. It utilizes only the local agent's information for feedback control, including \(\chi_i\), \(\hat\chi_i\), and \(W_{k,i}\), without involving any information exchange among neighboring AUVs.
\end{remark}
The following theorem summarizes the stability and tracking control performance results of the overall system:
\begin{theorem}
\label{theorem2}
Consider the local closed-loop system \eqref{eq:closedloopdynamics}. For each \(i \in \mathbf{I}[1, N]\), if there exists a sufficiently large compact set \(\Omega_{Z_i}\) such that \(Z_i \in \Omega_{Z_i}\) for all \(t \geq 0\), then for any bounded initial conditions, we have:
1) All signals in the closed-loop system remain uniformly ultimately bounded (UUB).
2) The position tracking error \(\eta_i - \eta_{d,i}\) converges exponentially to a small neighborhood around zero in finite time \(T_i > 0\) by choosing the design parameters with sufficiently large \(\underline{\lambda}(K_{1,i}) > 0\) and \(\underline{\lambda}(K_{2,i}) > 2\overline{\lambda}(K_{1,i}) > 0\), and sufficiently small \(\sigma_{k,i} > 0\) for all \(i \in \mathbf{I}[1, N]\) and \(k \in \mathbf{I}[1, 3]\).
\end{theorem}
\textbf{Proof:}
1) Consider the following Lyapunov function candidate for the closed-loop system \eqref{eq:closedloopdynamics}:
\begin{equation}
\begin{aligned}
    V_i &= \frac{1}{2}z_{1,i}^T z_{1,i} + \frac{1}{2}z_{2,i}^T M_i z_{2,i} + \frac{1}{2} \sum_{k=1}^{3} \tilde{W}_{k,i}^T \Gamma_{k,i}^{-1} \tilde{W}_{k,i}.
\end{aligned}
\label{eq:lyapunove}
\end{equation}
Evaluating the derivative of \(V_i\) along the trajectory of \eqref{eq:closedloopdynamics} for all \(i \in \mathbf{I}[1, N]\) yields:
\begin{equation}
\begin{aligned}
    \dot{V}_i &= z_{1,i}^T(-K_{1,i}z_{1,i} + J_i(\eta_i)z_{2,i}) \\
    &\quad + z_{2,i}^T\left(-J_i^T(\eta_i)z_{1,i} - K_{2,i}z_{2,i} + \tilde {W}_{k,i}^{T}S_{k,i}(Z_{i}) - \epsilon_{k,i}(Z_i))\right) \\
    &\quad - \sum_{k=1}^{3} \tilde{W}_{k,i}^T \left(S_{k,i}(Z_i)z_{2k,i} + \sigma_{k,i} \hat{W}_{k,i}\right) \\
    &= -z_{1,i}^T K_{1,i} z_{1,i} - z_{2,i}^T K_{2,i} z_{2,i} - z_{2,i}^T \epsilon_{k,i}(Z_i) \\
    &\quad - \sum_{k=1}^{3} \sigma_{k,i} \tilde{W}_{k,i}^T \hat{W}_{k,i}, \quad \forall i \in \mathbf{I}[1, N].
\end{aligned}
\end{equation}
Choose \(K_{2,i} = K_{1,i} + K_{22,i}\) such that \(K_{1,i}, K_{22,i} \in \mathbb{S}_3^+\). Using the completion of squares, we have:
\begin{equation}
\begin{aligned}
    -\sigma_{k,i} \tilde{W}_{k,i}^T \hat{W}_{k,i} &\leq -\frac{\sigma_{k,i} \|\tilde{W}_{k,i}\|^2}{2} + \frac{\sigma_{k,i} \|W_{k,i}^*\|^2}{2}, \\
    -z_{2,i}^T K_{22,i} z_{2,i} - z_{2,i}^T \epsilon_i(Z_i) &\leq \frac{\epsilon_i^T(Z_i) \epsilon_i(Z_i)}{4 \underline{\lambda}(K_{22,i})} \leq \frac{\|\epsilon_i^*\|^2}{4 \underline{\lambda}(K_{22,i})}.
\end{aligned}
\end{equation}
where \(\epsilon_i^* = [\epsilon_{1,i}^*, \epsilon_{2,i}^*, \epsilon_{3,i}^*]^T\). Then, we obtain:
\begin{equation}
\begin{aligned}
    \dot{V}_i \leq &-z_{1,i}^T K_{1,i} z_{1,i} - z_{2,i}^T K_{1,i} z_{2,i} + \frac{\|\epsilon_i^*\|^2}{4 \underline{\lambda}(K_{22,i})} \\
    &+ \sum_{k=1}^{3} ( -\frac{\sigma_{k,i} \|\tilde{W}_{k,i}\|^2}{2} + \frac{\sigma_{k,i} \|W_{k,i}^*\|^2}{2} .
\end{aligned}
\end{equation}
It follows that \(\dot{V}_i\) is negative definite whenever:
\begin{equation}
\begin{aligned}
    \|z_{1,i}\| &> \frac{\|\epsilon_i^*\|}{2\sqrt{\underline{\lambda}(K_{1,i}) \underline{\lambda}(K_{22,i})}} + \sum_{k=1}^{3} \left( \sqrt{\frac{\sigma_{k,i}}{2 \underline{\lambda}(K_{1,i})}} \|W_{k,i}^*\| \right), \\
    \|z_{2,i}\| &> \frac{\|\epsilon_i^*\|}{2\sqrt{\underline{\lambda}(K_{1,i}) \underline{\lambda}(K_{22,i})}} + \sum_{k=1}^{3} \left( \sqrt{\frac{\sigma_{k,i}}{2 \underline{\lambda}(K_{1,i})}} \|W_{k,i}^*\| \right), \\
    \|\tilde{W}_{k,i}\| &> \frac{\|\epsilon_i^*\|}{2\sqrt{\sigma_{k,i} \underline{\lambda}(K_{22,i})}} + \sum_{k=1}^{3} \|W_{k,i}^*\| \mathrel{\mathop:}= \tilde{W}_{k,i}^*.
\end{aligned}
\end{equation}
For all \(i \in \mathbf{I}[1, N]\), \(\exists k \in \mathbf{I}[1, 3]\). This leads to the Uniformly Ultimately Bounded (UUF) behavior of the signals \(z_{1,i}\), \(z_{2,i}\), and \(\tilde{W}_{k,i}\) for all \(i \in \mathbf{I}[1, N]\) and \(k \in \mathbf{I}[1, 3]\). As a result, it can be easily verified that since \(\eta_di = \eta_i + d^*_i\) with \(\eta_i\) bounded (according totheorem\ref{theorem1} and assumption\ref{assump1}), \(\eta_i = z_{1,i} + \eta_i\) is bounded for all \(i \in \mathbf{I}[1, N]\). Similarly, the boundedness of \(\nu_i = z_{2,i} + \alpha_i\) can be confirmed by the fact that \(\alpha_i\) in \eqref{eq:virtualcontrolinput} is bounded.
In addition, \(W_{k,i} = \tilde{W}_{k,i} + W_{k,i}^*\) is also bounded for all \(i \in \mathbf{I}[1, N]\) and \(k \in \mathbf{I}[1, 3]\) because of the boundedness of \(\tilde{W}_{k,i}\) and \(W_{k,i}^*\). Moreover, in light of \eqref{eq:desiredvritualcontinput}, \(\dot{\alpha}_i\) is bounded as all the terms on the right-hand side of \eqref{eq:desiredvritualcontinput} are bounded. This leads to the boundedness of the control signal \(\tau_i\) in \eqref{eq:ddlfeedback} since the Gaussian function vector \(S_i^F(Z_i)\) is guaranteed to be bounded for any \(Z_i\). As such, all the signals in the closed-loop system remain UUB, which completes the proof of the first part.

2) For the second part, it will be shown that \(\eta_i\) will converge arbitrarily close to \(\eta_{di}\) in some finite time \(T_i > 0\) for all \(i \in \mathbf{I}[1, N]\). To this end, we consider the following Lyapunov function candidate for the dynamics of \(z_{1,i}\) and \(z_{2,i}\) in \eqref{eq:closedloopdynamics}:
\begin{equation}
\begin{aligned}
    V_{z,i} &= \frac{1}{2}z_{1,i}^T z_{1,i} + \frac{1}{2}z_{2,i}^T M_i z_{2,i}, \quad \forall i \in \mathbf{I}[1, N].
\end{aligned}
\label{eq;lyapunov2}
\end{equation}
The derivative of \(V_{z,i}\) is:
\begin{equation}
\begin{aligned}
    \dot{V}_{z,i} &= z_{1,i}^T \left(-K_{1,i}z_{1,i} + J_i(\eta_i)z_{2,i}\right) + z_{2,i}^T \left(-J_i^T(\eta_i)z_{1,i} - K_{2,i}z_{2,i} + \tilde{W}_{k,i}^T S_{k,i}(Z_i) - \epsilon_i(Z_i)\right) \\[2pt]
    &= -z_{1,i}^T K_{1,i} z_{1,i} - z_{2,i}^T K_{2,i} z_{2,i} + z_{2,i}^T \tilde{W}_{k,i}^T S_{k,i}(Z_i) - z_{2,i}^T \epsilon_{k,i}(Z_i), \quad \forall i \in \mathbf{I}[1, N].
\end{aligned}
\end{equation}
Similar to the proof of part one, we let \(K_{2,i} = K_{1,i} + 2K_{22,i}\) with \(K_{1,i}, K_{22,i} \in \mathbb{S}_3^+\). According to \cite{wang2009deterministic}, the Gaussian RBF NN regressor \(S_i^F(Z_i)\) is bounded by \(\|S_i^F(Z_i)\| \leq s_i^*\) for any \(Z_i\) and for all \(i \in \mathbf{I}[1, N]\) with some positive number \(s_i^* > 0\). Through completion of squares, we have:
\begin{equation}
\begin{aligned}
    -z_{2,i}^T K_{22,i} z_{2,i} + z_{2,i}^T \tilde{W}_i^T S_i^F(Z_i) &\leq \frac{\|\tilde{W}_i^*\|^2 s_i^{*2}}{4 \underline{\lambda}(K_{22,i})}, \\[2pt]
    -z_{2,i}^T K_{22,i} z_{2,i} - z_{2,i}^T \epsilon_i(Z_i) &\leq \frac{\|\epsilon_i^*\|^2}{4 \underline{\lambda}(K_{22,i})}.
\end{aligned}
\end{equation}
where \(\tilde{W}_i^* = [\tilde{W}_{1,i}^*, \tilde{W}_{2,i}^*, \tilde{W}_{3,i}^*]^T\). This leads to:
\begin{equation}
\begin{aligned}
    \dot{V}_{z,i} &\leq -z_{1,i}^T K_{1,i} z_{1,i} - z_{2,i}^T K_{1,i} z_{2,i} + \delta_i \\[2pt]
    &\leq -2\underline{\lambda}(K_{1,i})\left(\frac{1}{2}z_{1,i}^T z_{1,i} + \frac{1}{2\overline{\lambda}(M_i)}z_{2,i}^T M_i z_{2,i}\right) + \delta_i \\[2pt]
    &\leq -\rho_i V_{z,i} + \delta_i, \quad \forall i \in \mathbf{I}[1, N],
\end{aligned}
\label{30}
\end{equation}
where \(\rho_i = \min\{2\underline{\lambda}(K_{1,i}), 2\underline{\lambda}(K_{1,i})/\overline{\lambda}(M_i)\}\) and \(\delta_i = (\|\tilde{W}_i^*\|^2 s_i^{*2}/4 \underline{\lambda}(K_{22,i})) + (\|\epsilon_i^*\|^2/4 \underline{\lambda}(K_{22,i}))\), \(\forall i \in \mathbf{I}[1, N]\). Solving the inequality \eqref{30} yields:
\begin{equation}
\begin{aligned}
  0 \leq V_{z,i}(t) \leq V_{z,i}(0)\exp(-\rho_i t) + \frac{\delta_i}{\rho_i} 
\end{aligned}
\end{equation}
which together with \eqref{eq;lyapunov2} implies that:
\begin{equation}
\begin{aligned}
    \min \left\{1, \underline{\lambda}(M_i)\right\} \frac{1}{2} \left(\|z_{1,i}\|^2 + \|z_{2,i}\|^2\right) \leq V_{z,i}(0)\exp \left({-\rho_i t}\right) + \frac{\delta_i}{\rho_i},
    &\quad \forall t \geq 0, \; i \in \mathbf{I}[1, N].
\end{aligned}
\end{equation}
Also
\begin{equation}
\begin{aligned}
    \|z_{1,i}\|^2 + \|z_{2,i}\|^2 \leq \frac{2}{\min \left\{1, \underline{\lambda}(M_i)\right\}} V_{z,i}(0) \exp \left(-\rho_i t\right) 
    + \frac{2 \delta_i}{\rho_i \min \left\{1, \underline{\lambda}(M_i)\right\}}.
\end{aligned}
\end{equation}
Consequently, it is straightforward that given \(\bar{\delta}_i > \sqrt{{2\delta_i}/{\rho_i \min\{1, \underline{\lambda}(M_i)\}}}\), there exists a finite time \(T_i > 0\) for all \(i \in \mathbf{I}[1, N]\) such that for all \(t \geq T_i\), both \(z_{1,i}\) and \(z_{2,i}\) satisfy \(\|z_{1,i}(t)\| \leq \bar{\delta}_i\) and \(\|z_{2,i}(t)\| \leq \bar{\delta}_i\) \(\forall i \in \mathbf{I}[1, N]\), where \(\bar{\delta}_i\) can be made arbitrarily small by choosing sufficiently large \(\underline{\lambda}(K_{1,i}) > 0\) and \(\underline{\lambda}(K_{2,i}) > 2 \overline{\lambda}(K_{1,i}) > 0\) for all \(i \in \mathbf{I}[1, N]\). This ends the proof.

By integrating the outcomes of theorems 1 and 2, the following theorem is established, which can be presented without additional proof:
\begin{theorem}
    \label{theorem3}
 By Considering the multi-AUV system \eqref{eq:auvdynamics} and the virtual leader dynamics \eqref{eq:leaderdynamics} with the network communication topology \(G\) and under assumptions \ref{assump1} and \ref{assump2}, the objective 1) of Problem 1 (i.e., \(\eta_i\) converges to \(\eta_0 + d_i^*\) exponentially for all \(i \in \mathbf{I}[1, N]\)) can be achieved by using the cooperative observer \eqref{eqn:DAO}, \eqref{eq:adaptationlaw} and the DDL control law \eqref{eq:ddlfeedback} and  \eqref{eq:robustlaw} with all the design parameters satisfying the requirements in theorems \ref{theorem1} and \ref{theorem2}, respectively.
\end{theorem}

\begin{remark} 
    \label{remark5}
    With the proposed two-layer formation learning control architecture, inter-agent information exchange occurs solely in the first-layer DA observation. Only the observer's estimated information, and not the physical plant state information, needs to be shared among neighboring agents. Additionally, since no global information is required for the design of each local AUV control system, the proposed formation learning control protocol can be designed and implemented in a fully distributed manner.
\end{remark}
\begin{remark} 
    \label{remark5}
    It is important to note that the eigenvalue constraints on \(A_0\) in \ref{assump1} are not needed for cooperative observer estimation (as detailed in the section \eqref{sec:Adaptive} or for achieving formation tracking control performance (as discussed in this section). This indicates that formation tracking control can be attained for general reference trajectories, including both periodic paths and straight lines, provided they are bounded. However, these constraints will become necessary in the next section to ensure the accurate learning capability of the proposed method.
\end{remark}
\section{Accurate learning from Formation Control}
\label{sec:closedloopformation}
It is necessary to demonstrate the convergence of the RBF NN weights in Equations \eqref{eq:ddlfeedback} and \eqref{eq:robustlaw} to their optimal values for accurate learning and identification. The main result of this section is summarized in the following theorem.
\begin{theorem}
\label{theorem4} Consider the local closed-loop system \eqref{eq:closedloopdynamics} with assumptions \ref{assump1} and \ref{assump2}. For each \(i \in \mathbf{I}[1, N]\), if there exists a sufficiently large compact set \(\Omega_{Z_i}\) such that \(Z_i \in \Omega_{Z_i}\) for all \(t \geq 0\), then for any bounded initial conditions and \(W_{k,i}(0) = 0\) \(\forall i \in \mathbf{I}[1, N], k \in \mathbf{I}[1, 3]\), the local estimated neural weights \(W_{\zeta,k,i}\) converge to small neighborhoods of their optimal values \(W_{\zeta,k,i}^*\) along the periodic reference tracking orbit \(\phi_{\zeta,i}(Z_i(t))|_{t \geq T_i}\) (denoting the orbit of the NN input signal \(Z_i(t)\) starting from time \(T_i\)). This leads to locally accurate approximations of the nonlinear uncertain dynamics \(f_{k,i}(Z_i)\) \(\forall k \in \mathbf{I}[1, 3]\) in \eqref{eq:nonlinearfunc} being obtained by \(W_{k,i}^T S_{k,i}(Z_i)\), as well as by \(\bar{W}_{k,i}^T S_{k,i}(Z_i)\), where \(\forall i \in \mathbf{I}[1, N], k \in \mathbf{I}[1, 3]\).

\begin{equation}
    \begin{aligned}
     \bar{W}_{k,i} = \mathrm {mean}_{t\in [t_{a,i}, t_{b,i}]}\hat {W}_{k,i}(t)   
     \label{34}
    \end{aligned}
\end{equation}
Where \([t_{a,i}, t_{b,i}]\) (\(t_{b,i} > t_{a,i} > T_i\)) represents a time segment after the transient process.
\end{theorem}
\textbf{Proof:} From theorem \ref{theorem3}, we have shown that for all \(i \in \mathbf{I}[1, N]\), \(\eta_i\) will closely track the periodic signal \(\eta_{d,i} = \eta_0 + d_i^*\) in finite time \(T_i\). In addition, \eqref{eq:virtualcontrolinput} implies that \(\nu_i\) will also closely track the signal \(J_i^T(\eta_i) \dot{\eta}_0^i\) since both \(z_{1,i}\) and \(z_{2,i}\) will converge to a small neighborhood around zero according to \ref{theorem2}. Moreover, since \(\dot{\eta}_0^i\) will converge to \(\dot{\eta}_0\) according to \ref{theorem1}, and \ref{theorem2} \(J_i(\eta_i)\) is a bounded rotation matrix, \(\nu_i\) will also be a periodic signal after finite time \(T_i\), because \(\dot{\eta}_0\) is periodic under \ref{assump1}.
Consequently, since the RBF NN input \(Z_i(t) = \mathrm{col}\{\eta_i, \nu_i\}\) becomes a periodic signal for all \(t \leq T_i\), the PE condition of some internal closed-loop signals, i.e., the RBF NN regression subvector \(S_{\zeta,k,i}(Z_i)\) (\(\forall t \geq T_i\)), is satisfied according to Lemma \ref{lemma1}. It should be noted that the periodicity of \(Z_i(t)\) leads to the PE of the regression subvector \(S_{\zeta,k,i}(Z_i)\), but not necessarily the PE of the whole regression vector \(S_{k,i}(Z_i)\). Thus, we term this as a partial PE condition, and we will show the convergence of the associated local estimated neural weights \(W_{\zeta,k,i} \rightarrow W_{\zeta,k,i}^*\), rather than \(W_{k,i} \rightarrow W_{k,i}^*\).

Thus, to prove accurate convergence of local neural weights \(W_{\zeta,k,i}\) associated with the regression subvector \(S_{\zeta,k,i}(Z_i)\) under the satisfaction of the partial PE condition, we first rewrite the closed-loop dynamics of \(z_{1,i}\) and \(z_{2,i}\) along the periodic tracking orbit \(\phi_{\zeta,i}(Z_i(t))|_{t \geq T_i}\) by using the localization property of the Gaussian RBF NN:
\begin{equation}
\begin{aligned}
    \dot{z}_{1,i} &= -K_{1,i} z_{1,i} + J_i(\eta_i) z_{2,i}, \\[2pt]
    \dot{z}_{2,i} &= M_i^{-1} \Bigl( -W_{\zeta,i}^{*T} S_{\zeta,i}^F(Z_i) - \epsilon_{\zeta,i} - J_i^T(\eta_i) z_{1,i} - K_{2,i} z_{2,i} + \hat{W}_{\zeta,i}^T S_{\zeta,i}^F(Z_i) + \hat{W}_{\bar{\zeta},i}^T S_{\bar{\zeta},i}^F(Z_i) \Bigr) \\[2pt]
    &= M_i^{-1} ( -J_i^T(\eta_i) z_{1,i} - K_{2,i} z_{2,i} + \tilde{W}_{\zeta,i}^T S_{\zeta,i}^F(Z_i) - \epsilon'_{\zeta,i} ).
\end{aligned}
\end{equation}
where {\small \(F_i(Z_i) = W_{\zeta,i}^{*T} S_{\zeta,i}^F(Z_i) + \epsilon_{\zeta,i}\)} with {\small \( 
W_{\zeta,i}^{*T} S_{\zeta,i}^F(Z_i) = [W_{\zeta,1,i}^{*T} S_{\zeta,1,i}(Z_i), W_{\zeta,2,i}^{*T} S_{\zeta,2,i}(Z_i), W_{\zeta,3,i}^{*T} S_{\zeta,3,i}(Z_i)]^T\)} and {\small \(\epsilon_{\zeta,i} = [\epsilon_{\zeta,1,i}, \epsilon_{\zeta,2,i}, \epsilon_{\zeta,3,i}]^T\)} being the approximation error. Additionally, {\small \(W_{\zeta,i}^T S_{\zeta,i}^F(Z_i) + W_{\bar{\zeta},i}^T S_{\bar{\zeta},i}^F(Z_i) = W_i^T S_i^F(Z_i)\)} with subscripts {\small \(\zeta\)} and {\small \(\bar{\zeta}\)} denoting the regions close to and far away from the periodic trajectory {\small \(\phi_{\zeta,i}(Z_i(t))|_{t \geq T_i}\)}, respectively. According to \cite{wang2009deterministic}, {\small \(\|W_{\bar{\zeta},i}^T S_{\bar{\zeta},i}^F(Z_i)\|\)} is small, and the NN local approximation error {\small \(\epsilon'_{\zeta,i} = \epsilon_{\zeta,i} - W_{\bar{\zeta},i}^T S_{\bar{\zeta},i}^F(Z_i)\)} with {\small \(\|\epsilon'_{\zeta,i}\| = O(\|\epsilon_{\zeta,i}\|)\)} is also a small number. Thus, the overall closed-loop adaptive learning system can be described by:

\begin{equation}
\begin{aligned}
\left[ \begin{array}{c}
    \dot{z}_{1,i} \\[2pt]
    \dot{z}_{2,i} \\[2pt]
    \hline
    \dot{\tilde{W}}_{\zeta,1,i} \\
    \dot{\tilde{W}}_{\zeta,2,i} \\
    \dot{\tilde{W}}_{\zeta,3,i}
\end{array} \right] &= \left[ \begin{array}{c|c}
    \begin{array}{cc}
        -K_{1,i} & J_i(\eta_i) \\[4pt]
        -M_i^{-1} J_i^T(\eta_i) & -M_i^{-1} K_{2,i}
    \end{array} & \Xi_i \\ \hline
    \begin{array}{c@{\hspace{10pt}}c}
        \hspace{30pt}0 &\hspace{40pt} -\Gamma_{\zeta,1,i} S_{\zeta,1,i}(Z_i) \\
        \hspace{30pt}0 &\hspace{40pt} -\Gamma_{\zeta,2,i} S_{\zeta,2,i}(Z_i) \\
        \hspace{30pt}0 &\hspace{40pt} -\Gamma_{\zeta,3,i} S_{\zeta,3,i}(Z_i)
    \end{array} & 0
\end{array} \right] \times \left[ \begin{array}{c}
    z_{1,i} \\
    z_{2,i} \\
    \hline
    \tilde{W}_{\zeta,1,i} \\
    \tilde{W}_{\zeta,2,i} \\
    \tilde{W}_{\zeta,3,i}
\end{array} \right] + \left[ \begin{array}{c}
    0 \\
    -\epsilon'_{\zeta,i} \\
    \hline
    -\sigma_{i,1} \Gamma_{\zeta,1,i} \hat{W}_{\zeta,1,i} \\
    -\sigma_{i,2} \Gamma_{\zeta,2,i} \hat{W}_{\zeta,2,i} \\
    -\sigma_{i,3} \Gamma_{\zeta,3,i} \hat{W}_{\zeta,3,i}
\end{array} \right]
\end{aligned}
\label{36}
\end{equation}

and
\begin{equation}
\begin{aligned}
\begin{bmatrix}
    \dot{\tilde{W}}_{\bar{\zeta},i,1} \\[2pt]
    \dot{\tilde{W}}_{\bar{\zeta},i,2} \\[2pt]
    \dot{\tilde{W}}_{\bar{\zeta},i,3}
\end{bmatrix} &= \begin{bmatrix}
    -\Gamma_{\bar{\zeta},1,i} \left( S_{\bar{\zeta},1,i}(Z_i) z_{2,i} + \sigma_{i,1} \hat{W}_{\bar{\zeta},1,i} \right) \\[5pt]
    -\Gamma_{\bar{\zeta},2,i} \left( S_{\bar{\zeta},2,i}(Z_i) z_{2,i} + \sigma_{i,2} \hat{W}_{\bar{\zeta},2,i} \right) \\[5pt]
    -\Gamma_{\bar{\zeta},3,i} \left( S_{\bar{\zeta},3,i}(Z_i) z_{2,i} + \sigma_{i,3} \hat{W}_{\bar{\zeta},3,i} \right)
\end{bmatrix}.
\end{aligned}
\end{equation}
where
\begin{equation}
\begin{aligned}
\Xi_i &= \begin{bmatrix}
    0 \\[2pt]
    M_i^{-1} \begin{bmatrix}
        S_{\zeta,1,i}^T(Z_i) & 0 & 0 \\[2pt]
        0 & S_{\zeta,2,i}^T(Z_i) & 0 \\[2pt]
        0 & 0 & S_{\zeta,3,i}^T(Z_i)
    \end{bmatrix}
\end{bmatrix}.
\end{aligned}
\end{equation}
for all \(i \in \mathbf{I}[1, N]\). The exponential stability property of the nominal part of subsystem \eqref{36} has been well-studied in \cite{wang2009deterministic}, \cite{yuan2011persistency}, and \cite{yuan2012performance}, where it is stated that PE of \(S_{\zeta,k,i}(Z_i)\) will guarantee exponential convergence of \((z_{1,i}, z_{2,i}, \tilde{W}_{\zeta,k,i}) = 0\) for all \(i \in \mathbf{I}[1, N]\) and \(k \in \mathbf{I}[1, 3]\). Based on this, since \(\|\epsilon'_{\zeta,i}\| = O(\|\epsilon_{\zeta,i}\|) = O(\|\epsilon_i\|)\), and \(\sigma_{k,i} \Gamma_{\zeta,k,i} \hat{W}_{\zeta,k,i}\) can be made small by choosing sufficiently small \(\sigma_{k,i}\) for all \(i \in \mathbf{I}[1, N]\), \(k \in \mathbf{I}[1, 3]\), both the state error signals \((z_{1,i}, z_{2,i})\) and the local parameter error signals \(\tilde{W}_{\zeta,k,i}\) (\(\forall i \in \mathbf{I}[1, N], k \in \mathbf{I}[1, 3]\)) will converge exponentially to small neighborhoods of zero, with the sizes of the neighborhoods determined by the RBF NN ideal approximation error \(\epsilon_i\) as in \eqref{15} and \(\sigma_{k,i} \|\Gamma_{\zeta,k,i} \hat{W}_{\zeta,k,i}\|\).
The convergence of \(W_{\zeta,k,i} \rightarrow W_{\zeta,k,i}^*\) implies that along the periodic trajectory \(\phi_{\zeta,i}(Z_i(t))|_{t \geq T_i}\), we have
\begin{equation}
    \begin{aligned}
    f_{k,i}(Z_i) &= W_{\zeta,k,i}^{*T} S_{\zeta,k,i}(Z_i) + \epsilon_{\zeta,k,i} \\[2pt]
    &= \hat{W}_{\zeta,k,i}^T S_{\zeta,k,i}(Z_i) - \tilde{W}_{\zeta,k,i}^T S_{\zeta,k,i}(Z_i) + \epsilon_{\zeta,k,i} \\[2pt]
    &= \hat{W}_{\zeta,k,i}^T S_{\zeta,k,i}(Z_i) + \epsilon_{\zeta_{1},k,i} \\[2pt]
    &= \bar{W}_{\zeta,k,i}^T S_{\zeta,k,i}(Z_i) + \epsilon_{\zeta_{2},k,i},
\end{aligned}
\end{equation}
where for all \(i \in \mathbf{I}[1, N]\), \(k \in \mathbf{I}[1, 3]\), \(\epsilon_{\zeta_{1},k,i} = \epsilon_{\zeta,k,i} - \tilde{W}_{\zeta,k,i}^T S_{\zeta,k,i}(Z_i) = O(\|\epsilon_{\zeta,i}\|)\) due to the convergence of \(\tilde{W}_{\zeta,k,i} \rightarrow 0\). The last equality is obtained according to the definition of \eqref{34} with \(\bar{W}_{\zeta,k,i}\) being the corresponding subvector of \(\bar{W}_{k,i}\) along the periodic trajectory \(\phi_{\zeta,i}(Z_i(t))|_{t \geq T_i}\), and \(\epsilon_{\zeta_{2},k,i}\) being an approximation error using \(\bar{W}_{\zeta,k,i}^T S_{\zeta,k,i}(Z_i)\). Apparently, after the transient process, we will have \(\epsilon_{\zeta_{2},k,i} = O(\epsilon_{\zeta_{1},k,i})\), \(\forall i \in \mathbf{I}[1, N]\), \(k \in \mathbf{I}[1, 3]\).
Conversely, for the neurons whose centers are distant from the trajectory \(\phi_{\zeta,i}(Z_i(t))|_{t \geq T_i}\), the values of \(\|S_{\bar{\zeta},k,i}(Z_i)\|\) will be very small due to the localization property of Gaussian RBF NNs. From the adaptation law (17) with \(W^k_i(0) = 0\), it can be observed that these small values of \(S_{\bar{\zeta},k,i}(Z_i)\) will only minimally activate the adaptation of the associated neural weights \(W_{\bar{\zeta},k,i}\). As a result, both \(W_{\bar{\zeta},k,i}\) and \(W_{\bar{\zeta},k,i}^T S_{\bar{\zeta},k,i}(Z_i)\), as well as \(\bar{W}_{\bar{\zeta},k,i}\) and \(\bar{W}_{\bar{\zeta},k,i}^T S_{\bar{\zeta},k,i}(Z_i)\), will remain very small for all \(i \in \mathbf{I}[1, N]\), \(k \in \mathbf{I}[1, 3]\) along the periodic trajectory \(\phi_{\zeta,i}(Z_i(t))|_{t \geq T_i}\).
This indicates that the entire RBF NN \(W_{k,i}^T S_{k,i}(Z_i)\) and \(\bar{W}_{k,i}^T S_{k,i}(Z_i)\) can be used to accurately approximate the unknown function \(f_{k,i}(Z_i)\) locally along the periodic trajectory \(\phi_{\zeta,i}(Z_i(t))|_{t \geq T_i}\), meaning that
\begin{align}
    f_{k,i}(Z_i) &= \hat{W}_{\zeta,k,i}^T S_{\zeta,k,i}(Z_i) + \epsilon_{\zeta_{1},k,i} = \hat{W}_{k,i}^T S_{k,i}(Z_i) + \epsilon_{1,k,i} \\[2pt]
    &= \bar{W}_{\zeta,k,i}^T S_{\zeta,k,i}(Z_i) + \epsilon_{\zeta_{2},k,i} = \bar{W}_{k,i}^T S_{k,i}(Z_i) + \epsilon_{2,k,i}.
\end{align}
with the approximation accuracy level of \(\epsilon_{1,k,i} = \epsilon_{\zeta_{1},k,i} - W_{\bar{\zeta},k,i}^T S_{\bar{\zeta},k,i}(Z_i) = O(\epsilon_{\zeta_{1},k,i}) = O(\epsilon_{k,i})\) and \(\epsilon_{2,k,i} = \epsilon_{\zeta_{2},k,i} - \bar{W}_{\bar{\zeta},k,i}^T S_{\bar{\zeta},k,i}(Z_i) = O(\epsilon_{\zeta_{2},k,i}) = O(\epsilon_{k,i})\) for all \(i \in \mathbf{I}[1, N]\), \(k \in \mathbf{I}[1, 3]\). This ends the proof.

\begin{remark}
    \label{remark7}
    The key idea in the proof of theorem \ref{theorem4} is inspired by \cite{wang2009deterministic}. For more detailed analysis on the learning performance, including quantitative analysis on the learning accuracy levels \(\epsilon_{1,i,k}\) and \(\epsilon_{2,i,k}\) as well as the learning speed, please refer to \cite{yuan2011persistency}. Furthermore, the AUV nonlinear dynamics \eqref{eq:nonlinearfunc} to be identified do not contain any time-varying random disturbances. This is important to ensure accurate identification/learning performance under the deterministic learning framework. To understand the effects of time-varying external disturbances on deterministic learning performance, interested readers are referred to \cite{yuan2012performance} for more details.
\end{remark}
\begin{remark}
    \label{remark8}
    Based on \eqref{34}, to obtain the constant RBF NN weights \(\bar{W}_{k,i}\) for all \(i \in \mathbf{I}[1, N]\), \(k \in \mathbf{I}[1, 3]\), one needs to implement the formation learning control law \eqref{eq:ddlfeedback}, \eqref{eq:robustlaw} first. Then, according to Theorem 4, after a finite-time transient process, the RBF NN weights \(W_{k,i}\) will converge to constant steady-state values. Thus, one can select a time segment \([t_{a,i}, t_{b,i}]\) with \(t_{b,i} > t_{a,i} > T_i\) for all \(i \in \mathbf{I}[1, N]\) to record and store the RBF NN weights \(W_{k,i}(t)\) for \(t \in [t_{a,i}, t_{b,i}]\). Finally, based on these recorded data, \(\bar{W}_{k,i}\) can be calculated off-line using \eqref{34}.
\end{remark}
\begin{remark}
    \label{remark9}
    It is shown in theorem \ref{theorem4} that locally accurate learning of each individual AUV's nonlinear uncertain dynamics can be achieved using localized RBF NNs along the periodic trajectory \(\phi_{\zeta,i}(Z_i(t))|_{t \geq T_i}\). The learned knowledge can be further represented and stored in a time-invariant fashion using constant RBF NNs, i.e., \(\bar{W}_{k,i}^T S_{k,i}(Z_i)\) for all \(i \in \mathbf{I}[1, N]\), \(k \in \mathbf{I}[1, 3]\). In contrast to many existing techniques (e.g., \cite{peng2017distributed} and \cite{peng2015containment}), this is the first time, to the authors’ best knowledge, that locally accurate identification and knowledge representation using constant RBF NNs are accomplished and rigorously analyzed for multi-AUV formation control under complete uncertain dynamics.
\end{remark}
\section{Formation Control with Pre-learned Dynamics}
\label{sec:formationprelearnd}
In this section, we will further address objective 2 of problem \ref{problem1}, which involves achieving formation control without readapting to the AUV's nonlinear uncertain dynamics. To this end, consider the multiple AUV systems \eqref{eq:auvdynamics} and the virtual leader dynamics \eqref{eq:leaderdynamics}. We employ the estimator observer \eqref{eqn:DAO}, \eqref{eq:adaptationlaw} to cooperatively estimate the leader's state information. Instead of using the DDL feedback control law \eqref{eq:ddlfeedback}, and self-adaptation law \eqref{eq:adaptationlaw}, we introduce the following constant RBF NN controller, which does not require online adaptation of the NN weights:
\begin{equation}
\begin{aligned}
    \tau_{i} = -J_{i}^{T}(\eta_{i})z_{1,i} - K_{2,i}z_{2,i} + \bar{W}_{i}^{T}S_{i}^{F}(Z_{i})
    \label{42}
\end{aligned}
\end{equation}
where \(\bar{W}_{ki}^{T}S_{ki}^{F}(Z_{i}) = [\bar{W}_{1,i}^{T}S_{1,i}(Z_{i}), \bar{W}_{2,i}^{T}S_{2,i}(Z_{i}), \bar{W}_{3,i}^{T}S_{3,i}(Z_{i})]^T\) is ubtained from \eqref{34}. The term \(\bar{W}_{k,i}^T S_{k,i}(Z_{i})\) represents the locally accurate RBF NN approximation of the nonlinear uncertain function \(f_{k,i}(Z_{i})\) along the trajectory \(\phi_{\zeta,i}(Z_{i}(t))|_{t \geq T_i}\), and the associated constant neural weights \(\bar{W}_{k,i}\) are obtained from the formation learning control process as discussed in remark \ref{remark8}.
\begin{theorem}
    \label{theorem5}
    Consider the multi-AUV system \eqref{eq:auvdynamics} and the virtual leader dynamics \eqref{eq:leaderdynamics} with the network communication topology \(G\). Under assumptions \ref{assump1} and \ref{assump2}, the formation control performance (i.e., \(\eta_i\) converges to \(\eta_0 + d^*_i\) exponentially with the same \(\eta_0\) and \(d^*_i\) defined in theorem \ref{theorem3} for all \(i \in \mathbf{I}[1, N]\)) can be achieved by using the DA observer \eqref{eqn:DAO}, \eqref{42} and the constant RBF NN control law \eqref{eq:adaptationlaw} with the constant NN weights obtained from \eqref{34}.
\end{theorem}
\textbf{Proof:} The closed-loop system for each local AUV agent can be established by integrating the controller \eqref{42} with the AUV dynamics \eqref{eq:auvdynamics}. 
\begin{equation}
    \begin{aligned}
       \dot {z}_{1,i}=&-K_{1,i}z_{1,i} + J_{i}(\eta _{i})z_{2,i} \\
       \dot {z}_{2,i}=&M_{i}^{-1}(-J_{i}^{T}(\eta _{i})z_{1,i} - K_{2,i}z_{2,i} + \bar {W}_{i}^{T}S_{i}^{F}(Z_{i}) - F_{i}(Z_{i}))\\
       =&M_{i}^{-1}(-J_{i}^{T}(\eta _{i})z_{1,i} - K_{2,i}z_{2,i} - \epsilon _{2,i}), \quad \forall i\in \mathbf {I}\left [{1,N}\right ]  
    \end{aligned}
\end{equation}
where \(\epsilon_{2,i} = [\epsilon_{21,i}, \epsilon_{22,i}, \epsilon_{23,i}]^T\). Consider the Lyapunov function candidate \(V_{z,i} = \frac{1}{2} z_{1,i}^T z_{1,i} + \frac{1}{2} z_{2,i}^T M_i z_{2,i}\), whose derivative along the closed-loop system described is given by:
\begin{equation}
    \begin{aligned}
        \dot{V}_{z,i} =& z_{1,i}^T \left(-K_{1,i} z_{1,i} + J_i(\eta_i) z_{2,i}\right) + z_{2,i}^T \left(-J_i^T(\eta_i) z_{1,i} - K_{2,i} z_{2,i} - \epsilon_{2,i}\right) \\
        =& -z_{1,i}^T K_{1,i} z_{1,i} - z_{2,i}^T K_{2,i} z_{2,i} - z_{2,i}^T \epsilon_{2,i}.
    \end{aligned}
\end{equation}
Selecting \(K_{2,i} = K_{1,i} + K_{22,i}\) where \(K_{1,i}, K_{22,i} \in \mathbb{S}_3^+\), we can utilize the method of completing squares to obtain:
\begin{equation}
    -z_{2,i}^T K_{22,i} z_{2,i} - z_{2,i}^T \epsilon_{2,i} \leq \left(\frac{\|\epsilon_{2,i}\|^2}{4 \underline{\lambda}(K_{22,i})}\right) \leq \left(\frac{\|\epsilon^*_{2,i}\|^2}{4 \underline{\lambda}(K_{22,i})}\right),
\end{equation}
which implies that:
\begin{equation}
    \dot{V}_{z,i} \leq -z_{1,i}^T K_{1,i} z_{1,i} - z_{2,i}^T K_{1,i} z_{2,i} + \left(\frac{\|\epsilon^*_{2,i}\|^2}{4 \underline{\lambda}(K_{22,i})}\right) \leq -\rho_i V_{z,i} + \delta_i, \quad \forall i\in \mathbf {I}\left [{1,N}\right ]
\end{equation}
where \(\rho_i = \min\{2 \underline{\lambda}(K_{1,i}), (2 \underline{\lambda}(K_{1,i}) / \overline{\lambda}(M_i))\}\) and \(\delta_i = (\|\epsilon^*_{2,i}\|^2 / 4 \underline{\lambda}(K_{22,i}))\).
Using similar reasoning to that in the proof of theorem \ref{theorem2}, it is evident from the derived inequality that all signals within the closed-loop system remain bounded. Additionally, \(\eta_i - \eta^d_i\) will converge to a small neighborhood around zero within a finite period. The magnitude of this neighborhood can be minimized by appropriately choosing large values for \(\underline{\lambda}(K_{1,i}) > 0\) and \(\underline{\lambda}(K_{2,i}) > \overline{\lambda}(K_{1,i})\) across all \(i \in \mathbf{I}[1, N]\). In line with theorem \ref{theorem1}, under assumptions \ref{assump1} and \ref{assump2}, the implementation of the DA observer  DA observer \eqref{eqn:DAO}, \eqref{eq:adaptationlaw} facilitates the exponential convergence of \(\eta_i\) towards \(\eta_0\). This conjunction of factors assures that \(\eta_i\) rapidly aligns with \(\eta_{d,i} = \eta_0 + d^*_i\), achieving the objectives set out for formation control.
\begin{remark}
    \label{remark10}
    Building on the locally accurate learning outcomes discussed in Section~\ref{sec:closedloopformation}, the newly developed distributed control protocol comprising \eqref{eqn:DAO}, \eqref{eq:adaptationlaw}, and  \eqref{42} facilitates stable formation control across a repeated formation pattern. Unlike the formation learning control approach outlined in Section~\ref{SecondLayer}, which involves \eqref{eqn:DAO}, \eqref{eq:adaptationlaw} coupled with \eqref{eq:ddlfeedback} and \eqref{eq:robustlaw}, the current method eliminates the need for online RBF NN adaptation for all AUV agents. This significantly reduces the computational demands, thereby enhancing the practicality of implementing the proposed distributed RBF NN formation control protocol. This innovation marks a significant advancement over many existing techniques in the field.
\end{remark}
\section{Simulation}
\label{sec:simulation}
We consider a multi-AUV heterogeneous system composed of 5 AUVs for the simulation. The dynamics of these AUVs are described in THE system \eqref{eq:auvdynamics}. The system parameters for each AUV are specified as follows:
\begin{align}
    \begin{aligned}
        M_{i} &= \begin{bmatrix} 
            m_{11,i} & 0 & 0 \\ 
            0 & m_{22,i} & m_{23,i} \\ 
            0 & m_{23,i} & m_{33,i} 
        \end{bmatrix},\\
        C_{i} &= \begin{bmatrix} 
            0 & 0 & -m_{22,i} v_{i} - m_{23,i} r_{i} \\ 
            0 & 0 & -m_{11,i} u_{i} \\ 
            m_{22,i} v_{i} + m_{23,i} r_{i} & -m_{11,i} u_{i} & 0 
        \end{bmatrix},\\
        D_{i} &= \begin{bmatrix} 
            d_{11,i} (\nu_{i}) & 0 & 0 \\ 
            0 & d_{22,i} (\nu_{i}) & d_{23,i} (\nu_{i}) \\ 
            0 & d_{32,i} (\nu_{i}) & d_{33,i} (\nu_{i}) 
        \end{bmatrix}, \quad g_{i} = 0,\\
        \Delta_{i} &= \begin{bmatrix} 
            \Delta_{1,i} (\chi_{i}) \\ 
            \Delta_{2,i} (\chi_{i}) \\ 
            \Delta_{3,i} (\chi_{i}) 
        \end{bmatrix}, \quad \forall i \in \mathbf{I} [1, 5]
    \end{aligned}
\end{align}
where the mass and damping matrix components for each AUV \( i \) are defined as:
\begin{align*}
    m_{11,i} &= m_i - X_{\dot{u},i}, & m_{22,i} &= m_i - Y_{\dot{v},i}, \\
    m_{23,i} &= m_i x_{g,i} - Y_{\dot{r},i}, & m_{33,i} &= I_{z,i} - N_{\dot{r},i}, \\
    d_{11,i} &= -(X_{u,i} + X_{uu,i} \|u_i\|), & d_{22,i} &= -(Y_{v,i} + Y_{vv,i} \|v_i\| + Y_{rv,i} \|r_i\|), \\
    d_{23,i} &= -(Y_{r,i} + Y_{vr,i} \|v_i\| + Y_{rr,i} \|r_i\|), & d_{32,i} &= -(N_{v,i} + N_{vv,i} \|v_i\| + N_{rv,i} \|r_i\|), \\
    d_{33,i} &= -(N_{r,i} + N_{vr,i} \|v_i\| + N_{rr,i} \|r_i\|).
\end{align*}
 according to the notations in \cite{prestero2001verification} and \cite{skjetne2005adaptive} the coefficients \(\{X(\cdot), Y(\cdot), N(\cdot)\}\) are hydrodynamic parameters. For the associated system parameters are borrowed from \cite{skjetne2005adaptive} (with slight modifications for different AUV agents) and simulation purposes and listed in table \ref{tabel1}. For all \( i \in \mathbf{I}[1,5] \), we set \( x_{g,i} = 0.05 \) and \( Y_{\dot{r},i} = Y_{rv,i} = Y_{vr,i} = Y_{rr,i} = N_{rv,i} = N_{rr,i} = N_{vv,i} = N_{vr,i} = N_{r,i} = 0 \).
Model uncertainties are given by:
\begin{table}[ht]
\centering
\begin{tabular}{l c c c c c} 
\toprule
\textbf{Parameter} & \textbf{AUV 1} & \textbf{AUV 2} & \textbf{AUV 3} & \textbf{AUV 4} & \textbf{AUV 5} \\
\midrule
$m$ (kg) & 23 & 25 & 20 & 30 & 35 \\
$I_z$ (kg$\cdot$m$^2$) & 1.8 & 2.0 & 1.5 & 2.2 & 2.5 \\
$X_{\dot{u}}$ (kg) & -2.0 & -2.5 & -1.5 & -2.5 & -3.0 \\
$Y_{\dot{v}}$ (kg) & -10 & -10 & -10 & -15 & -15 \\
$N_{\dot{r}}$ (kg$\cdot$m$^2$) & -1.0 & -1.5 & -1.0 & -2.5 & -2.5 \\
$X_u$ (kg/s) & -0.8 & -1.0 & -1.0 & -1.5 & -2.0 \\
$Y_v$ (kg/s) & -0.9 & -1.0 & -0.8 & -1.5 & -1.5 \\
$Y_r$ (kg/s) & 0.1 & 0.2 & 0.1 & 0.2 & 0.5 \\
$N_v$ (kg$\cdot$m$^2$/s) & 0.1 & 0.1 & 0.05 & 0.3 & 0.35 \\
$X_{uu}$ (kg/m) & -1.3 & -1.3 & -1.0 & -0.85 & -1.5 \\
$Y_{vv}$ (kg/m) & -36 & -25 & -20 & -15 & -20 \\
\bottomrule
\end{tabular}
\caption{Parameters of AUVs}
\label{tabel1}
\end{table}
\begin{align*}
    \Delta _{1}=&0, \quad \Delta _{2} =
    \begin{bmatrix} 0.2u_{2}^{2}+0.3v_{2} & -0.95 & 0.33\|r_{2}\| \end{bmatrix}^{T}\\
    \Delta _{3}=&\begin{bmatrix} -0.58+\cos \left ({v_{3}}\right ) & 0.23r_{3}^{3} & 0.74u_{3}^{2} \end{bmatrix}^{T}\\
    \Delta _{4}=&\begin{bmatrix} -0.31 & 0 & 0.38u_{4}^{2} + v_{4}^{3} \end{bmatrix}^{T}\\ \Delta _{5}=&\begin{bmatrix} \sin \left ({v_{5}}\right ) & \cos \left ({u_{5} + r_{5}}\right ) & -0.65 \end{bmatrix}^{T}. 
\end{align*}
Fig.~\ref{fig:topology} illustrates the communication topology and the spanning tree where agent 0 is the virtual leader and is considered as the root, in accordance with assumption~\ref{assump2}. The desired formation pattern requires each AUV, \( \eta_i \), to track a periodic signal generated by the virtual leader \( \eta_0 \). The dynamics of the leader are defined as follows:
\begin{figure}[h!]
\begin{center}
\includegraphics[width=10cm]{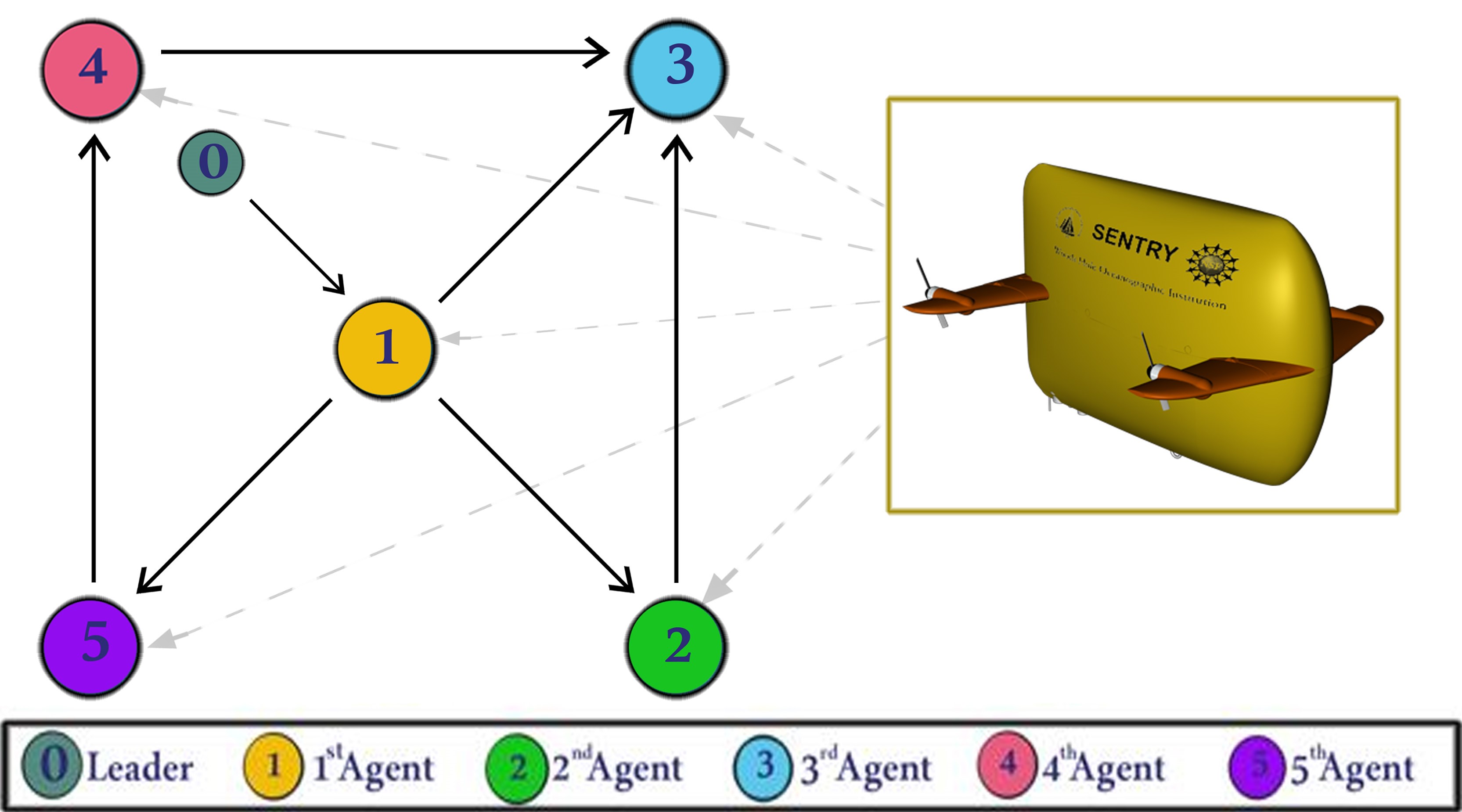}
\end{center}
\caption{Network topology of multi-AUV system with 0 as virtual leader}
\label{fig:topology}
\end{figure}
\begin{equation}
    \begin{aligned}
        \begin{bmatrix} \dot {\eta }_{0} \\ \dot {\nu }_{0} \end{bmatrix} = & \begin{bmatrix} 0 & \begin{bmatrix} 1 & 0 & 0 \\ 0 & -1 & 0 \\ 0 & 0 & 1 \end{bmatrix} \\ \begin{bmatrix} -1 & 0 & 0 \\ 0 & 1 & 0 \\ 0 & 0 & -1 \end{bmatrix} & 0 \end{bmatrix} \begin{bmatrix} {\eta }_{0} \\ {\nu }_{0} \end{bmatrix}, \\
        \begin{bmatrix} {\eta }_{0}(0) \\ {\nu }_{0}(0) \end{bmatrix} = & \begin{bmatrix} 0 & 80 & 0 & 80 & 0 & 80 \end{bmatrix}^{T}.
    \end{aligned}
    \label{eqn:leaderdynamicwithinitials}
\end{equation}
The initial conditions and system matrix are structured to ensure all eigenvalues of \( A_0 \) lie on the imaginary axis, thus satisfying Assumption~\ref{assump1}. The reference trajectory for \( \eta_0 \) is defined as \( [80\sin(t), 80\cos(t), 80\sin(t)]^T \). The predefined offsets \( d^*_i \), which determine the relative positions of the AUVs to the leader, are specified as follows:
\begin{align*}
    d^*_1 = [0, 0, 0]^T,\quad d^*_4 = [-10, 10, 0]^T,\quad d^*_2 = [10, -10, 0]^T,\quad d^*_5 = [-10, -10, 0]^T,\quad d^*_3 = [10, 10, 0]^T
\end{align*}
Each AUV tracks its respective position in the formation by adjusting its location to \( \eta_i = \eta_0 + d^*_i \).
\subsection{DDL Formation Learning Control Simulation}
\label{sec:DDLsimulation}
The estimated virtual leader's state, derived from the cooperative estimator in the first layer (see equation \eqref{eqn:DAO} and \eqref{eq:adaptationlaw}), is utilized to estimate each agent's complete uncertain dynamics within the DDL controller (second layer) using equations \eqref{eq:ddlfeedback} and \eqref{eq:robustlaw}. The uncertain nonlinear functions $F_i(Z_i)$ for each agent are approximated using RBF NNs, as described in equation \eqref{eq:nonlinearfunc}. Specifically, for each agent $i \in \{1, \dots, 5\}$, the nonlinear uncertain functions $F_i(Z_i)$, dependent on $\nu_i$, are modeled. The input to the NN, $Z_i = [u_i, v_i, r_i]^T$, allows the construction of Gaussian RBF NNs, represented by $W^T_{k,i} S_{k,i}(Z_i)$, utilizing $4096$ neurons arranged in an $16 \times 16 \times 16$ grid. The centers of these neurons are evenly distributed over the state space $[-100, 100] \times [-100, 100] \times [-100, 100]$, and each has a width $\gamma_{k,i} = 60$, for all $i \in \{1, \dots, 5\}$ and $k \in \{1, 2, 3\}$.

The observer and controller parameters are chosen as $\beta_1 = \beta_2 = 5$, and the diagonal matrices $K_{1,i} = 800 * \text{diag}\{1.2, 1, 1\}$ and $K_{2,i} = 1200 * \text{diag}\{1.2, 1, 1\}$, with $\Gamma_{k,i} = 10$ and $\sigma_{k,i} = 0.0001$ for all $i \in \{1, \dots, 5\}$ and $k \in \{1, 2, 3\}$. The initial conditions for the agents are set as $\eta_1(0) = [30, 60, 0]^T$, $\eta_2(0) = [40, 70, 0]^T$, $\eta_3(0) = [50, 80, 0]^T$, $\eta_4(0) = [10, 70, 0]^T$, and $\eta_5(0) = [10, 50, 0]^T$. Zero initial conditions are assumed for all the distributed observer states $(\chi^i_0, A^i_0)$ and the DDL controller states $W^k_i$ for all $i \in \{1, \dots, 5\}$ and $k \in \{1, 2, 3\}$. Time-domain simulation is carried out using the DDL formation learning control laws as specified in equations \eqref{eq:ddlfeedback} and \eqref{eq:robustlaw}, along with equations \eqref{eqn:DAO} and \eqref{eq:adaptationlaw}.

Figure~\ref{fig:Cooperativeestimator} displays the simulation results of the cooperative estimator (first layer) for all five agents. It illustrates how each agent's estimated states,\(\hat{\eta}_i\), converge perfectly to the leader's states \(\hat{\eta}_0\) thorough \eqref{eqn:DAO} and \eqref{eq:adaptationlaw}.
\setcounter{figure}{3}
\setcounter{subfigure}{0}
\begin{subfigure}
\setcounter{figure}{3}
\setcounter{subfigure}{0}
    \centering
    \begin{minipage}[b]{0.5\textwidth}
        \includegraphics[width=\linewidth]{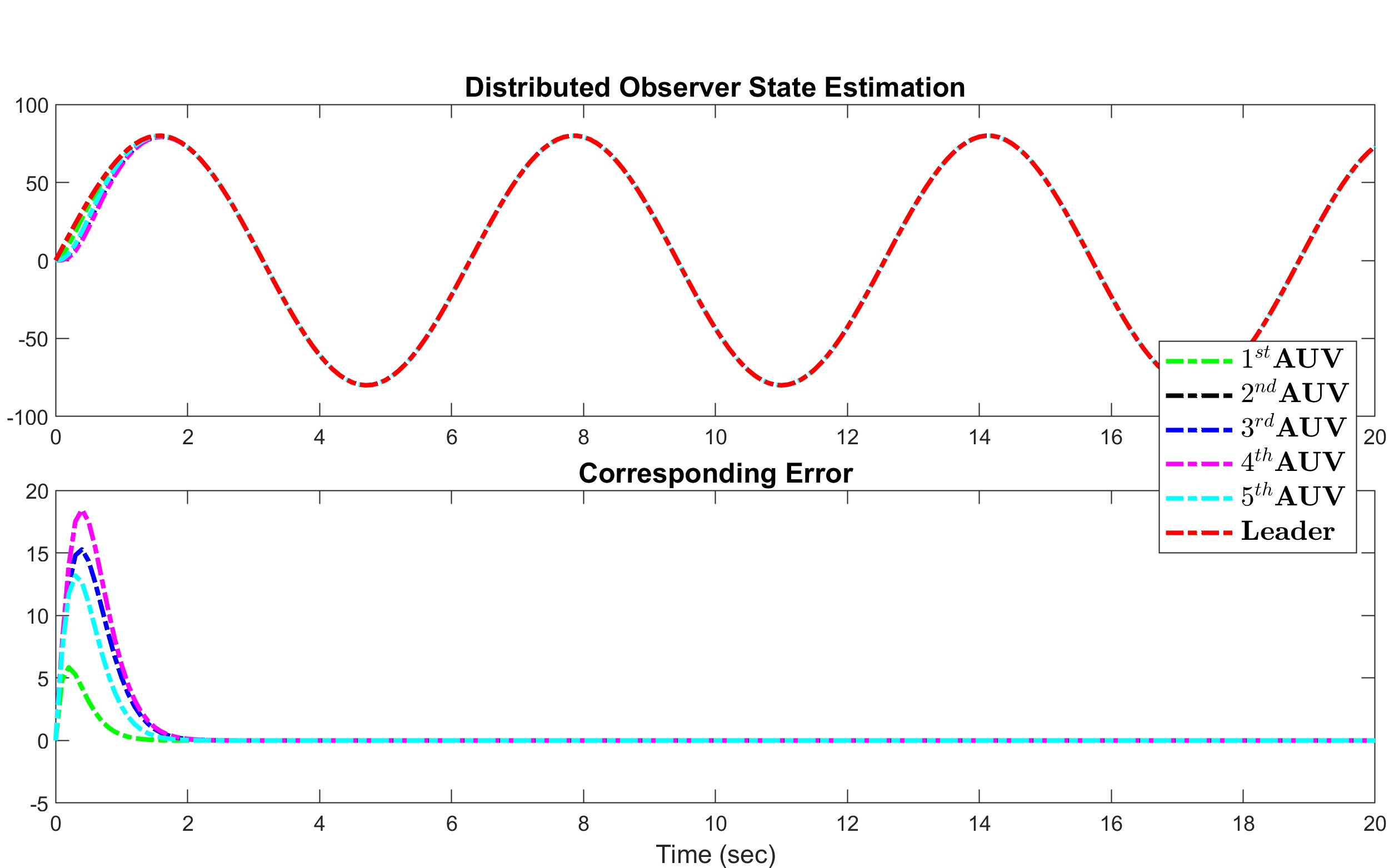}
        \caption{\(\hat{x}_i \rightarrow x_0 \, \text{(m)}\).}
        \label{fig:Subfigure 1}
    \end{minipage}  
   
\setcounter{figure}{3}
\setcounter{subfigure}{1}
    \begin{minipage}[b]{0.5\textwidth}
        \includegraphics[width=\linewidth]{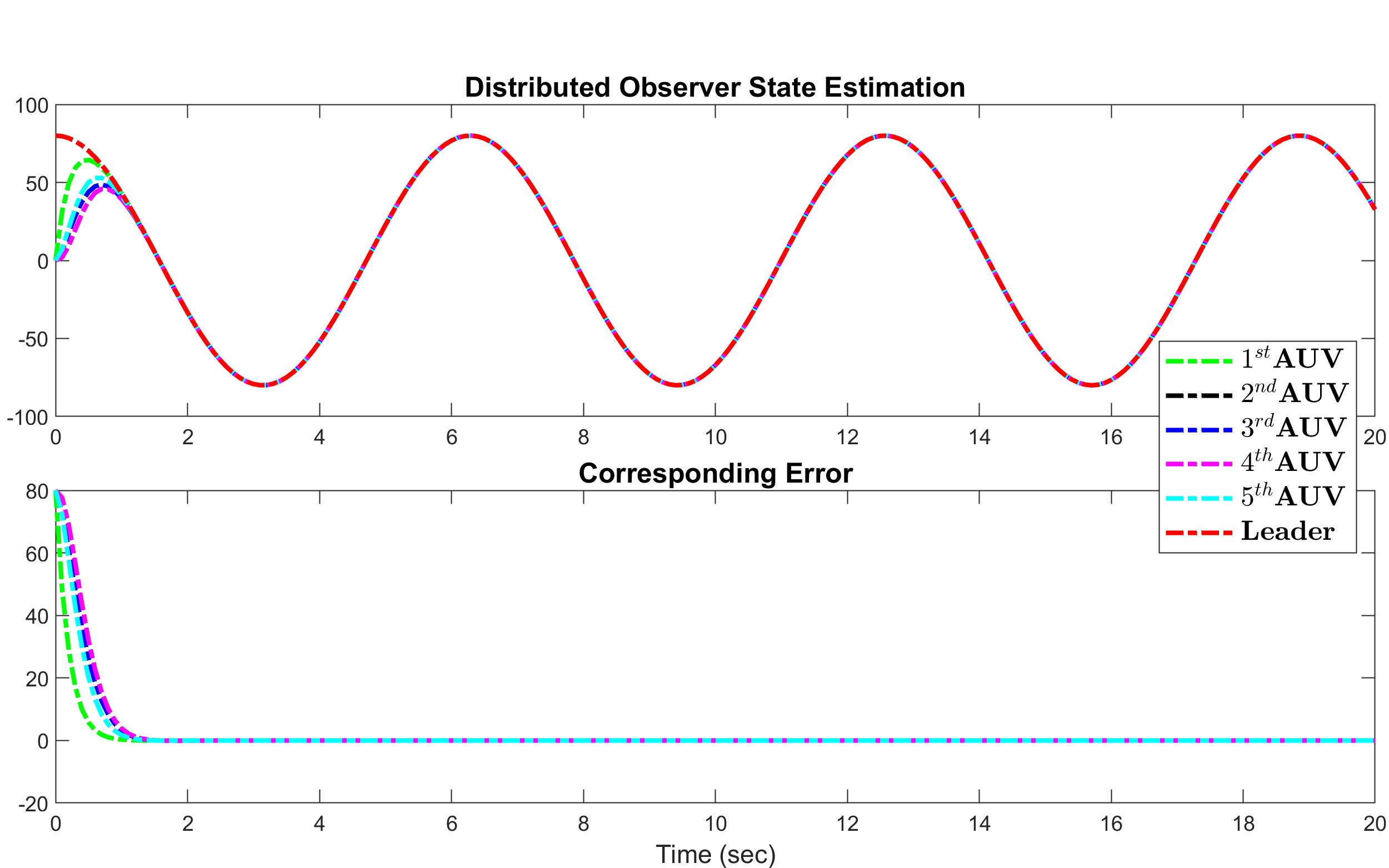}
        \caption{\(\hat{y}_i \rightarrow y_0 \, \text{(m)}\).}
        \label{fig:Subfigure 2}
    \end{minipage}

\setcounter{figure}{3}
\setcounter{subfigure}{2}
    \begin{minipage}[b]{0.5\textwidth}
        \includegraphics[width=\linewidth]{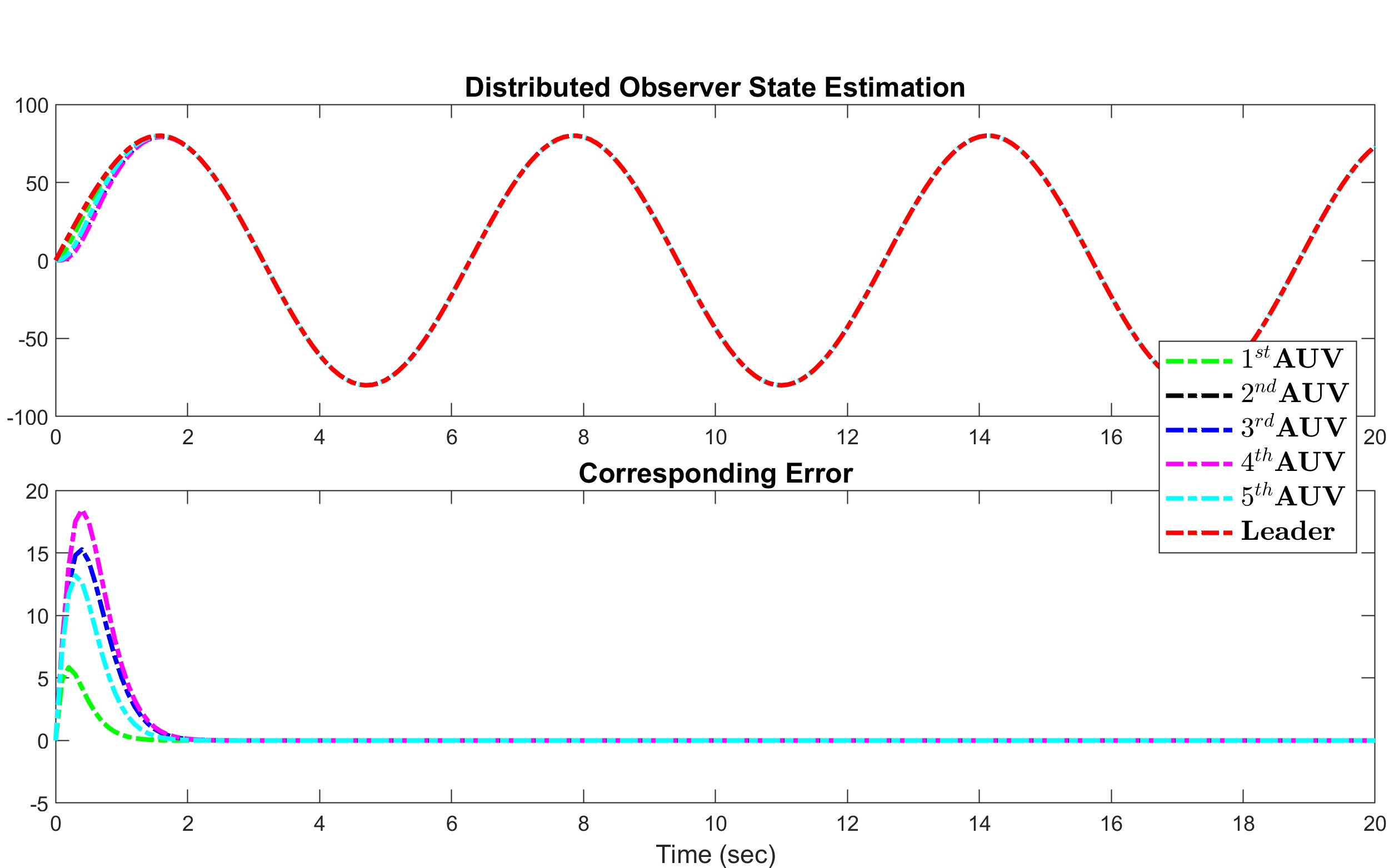}
        \caption{\(\hat\psi_i \rightarrow \psi_0 \, \text{(deg)}\).}
        \label{fig:Subfigure 3}
    \end{minipage}
    
\setcounter{figure}{3}
\setcounter{subfigure}{-1}
    \caption{Distributed observer for all  three states of each AUVs}
    \label{fig:Cooperativeestimator}
\end{subfigure}
Fig~\ref{fig:trackcont} presents the position tracking control responses of all agents. Fig~\ref{fig:trackcont1}a, \ref{fig:trackcont1}b, and \ref{fig:trackcont1}c illustrate the tracking performance of AUVs along the x-axis, y-axis, and vehicle heading, respectively, demonstrating effective tracking of the leader's position signal. While the first AUV exactly tracks the leader's states, agents 2 through 5 are shown to successfully follow agent 1, maintaining prescribed distances and alignment along the x and y axes, and matching the same heading angle. These results underscore the robustness of the real-time tracking control system, which enforces a predefined formation pattern, initially depicted in Fig~\ref{fig:topology}. Additionally, Fig~\ref{fig:formation} highlights the real-time control performance for all agents, showcasing the effectiveness of the tracking strategy in maintaining the formation pattern.
\setcounter{figure}{4}
\setcounter{subfigure}{0}
\begin{subfigure}
\setcounter{figure}{4}
\setcounter{subfigure}{0}
    \centering
    \begin{minipage}[b]{0.5\textwidth}
        \includegraphics[width=\linewidth]{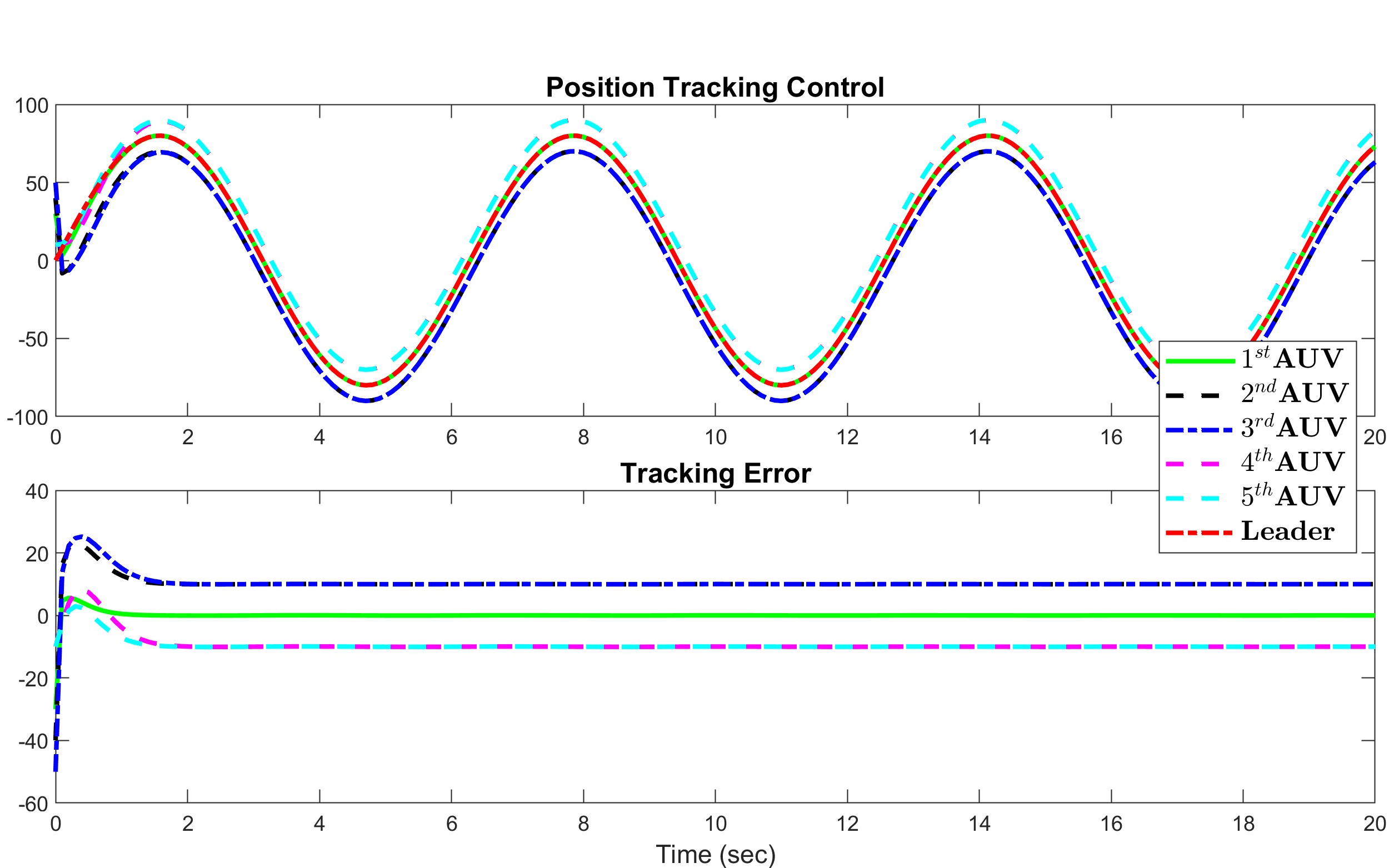}
        \caption{\(x_i \rightarrow x_0 \, \text{(m)}\).}
        \label{fig:trackcont1}
    \end{minipage}  
   
\setcounter{figure}{4}
\setcounter{subfigure}{1}
    \begin{minipage}[b]{0.5\textwidth}
        \includegraphics[width=\linewidth]{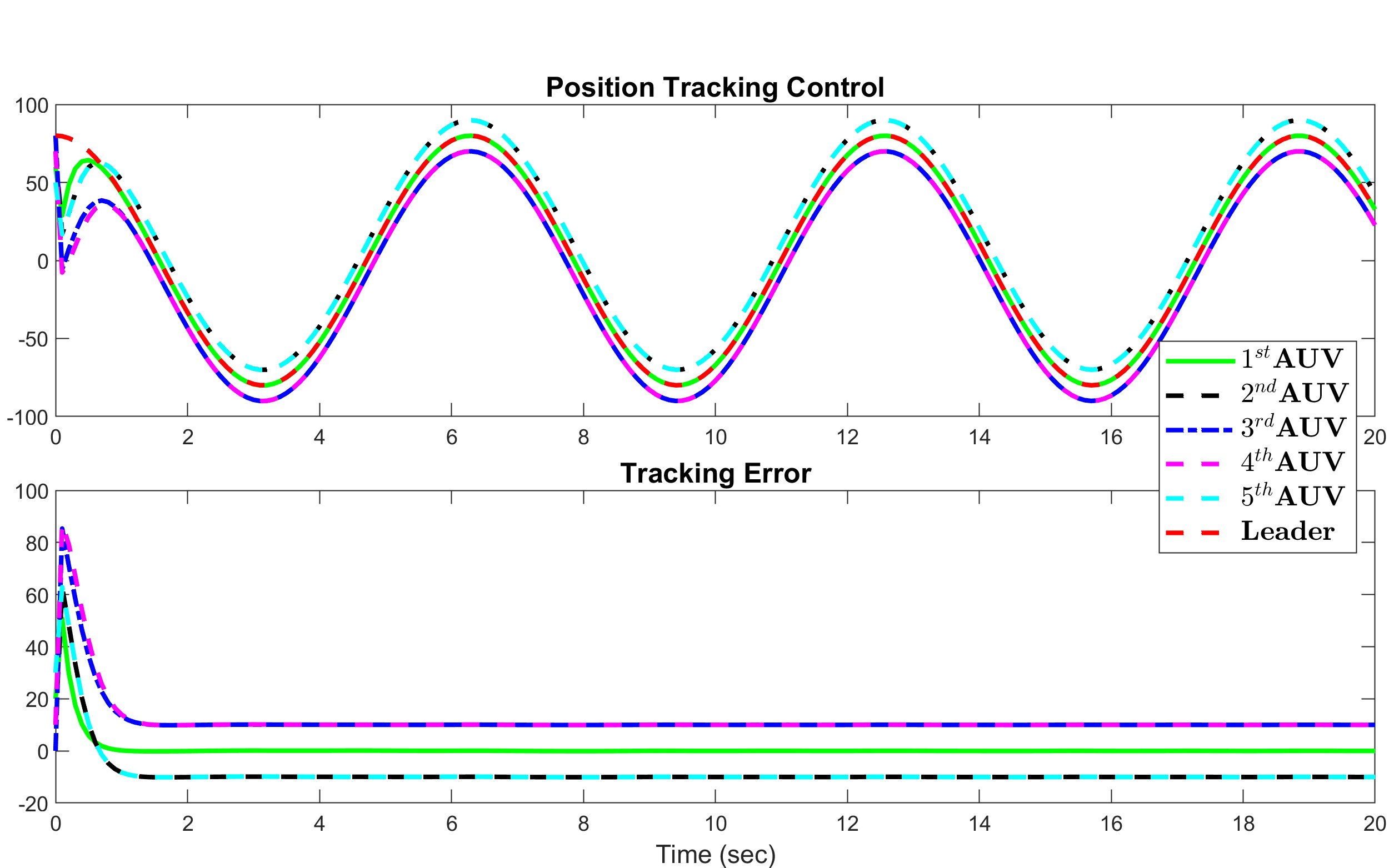}
        \caption{\(y_i \rightarrow y_0 \, \text{(m)}\).}
        \label{fig:trackcont2}
    \end{minipage}

\setcounter{figure}{4}
\setcounter{subfigure}{2}
    \begin{minipage}[b]{0.5\textwidth}
        \includegraphics[width=\linewidth]{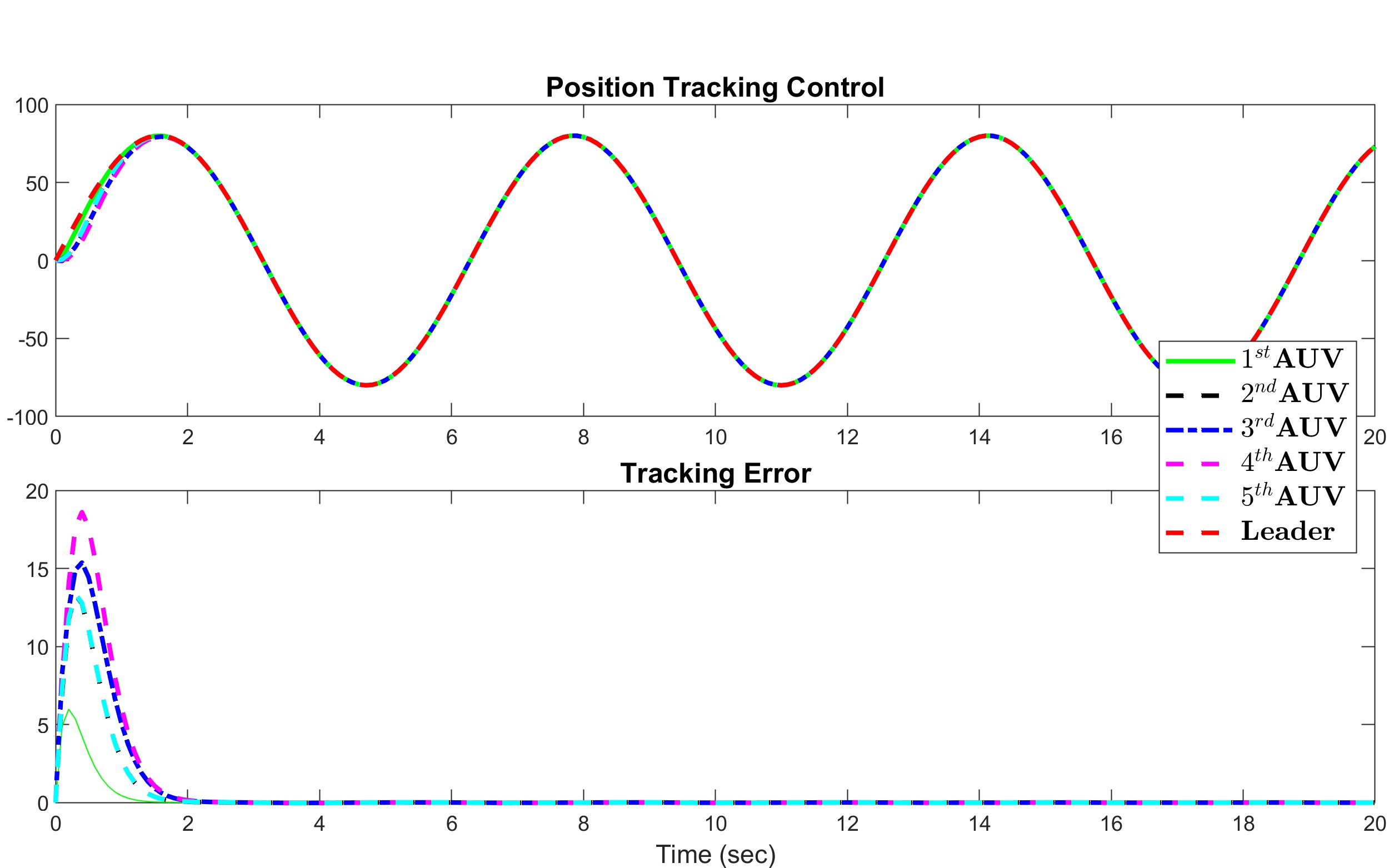}
        \caption{\(\psi_i \rightarrow \psi_0 \, \text{(deg)}\).}
        \label{fig:trackcont3}
    \end{minipage}
    
\setcounter{figure}{4}
\setcounter{subfigure}{-1}
    \caption{Postition Tracking Control for all three states of each AUVs}
    \label{fig:trackcont}
\end{subfigure}

\begin{figure}[h!]
\begin{center}
\includegraphics[width=10cm]{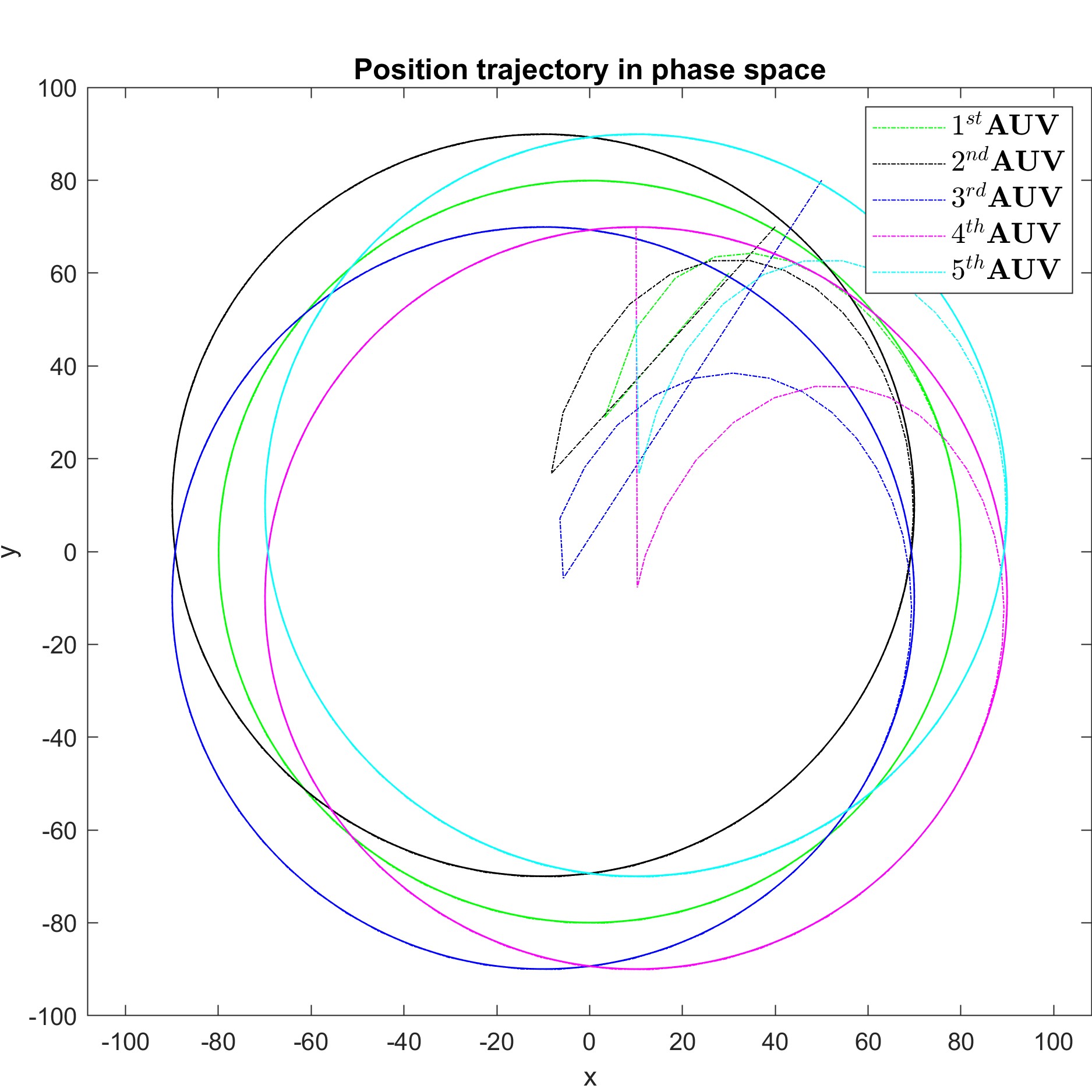}
\end{center}
\caption{Fomation control for all agents}
\label{fig:formation}
\end{figure}
\begin{figure}[h!]
\begin{center}
\includegraphics[width=10cm]{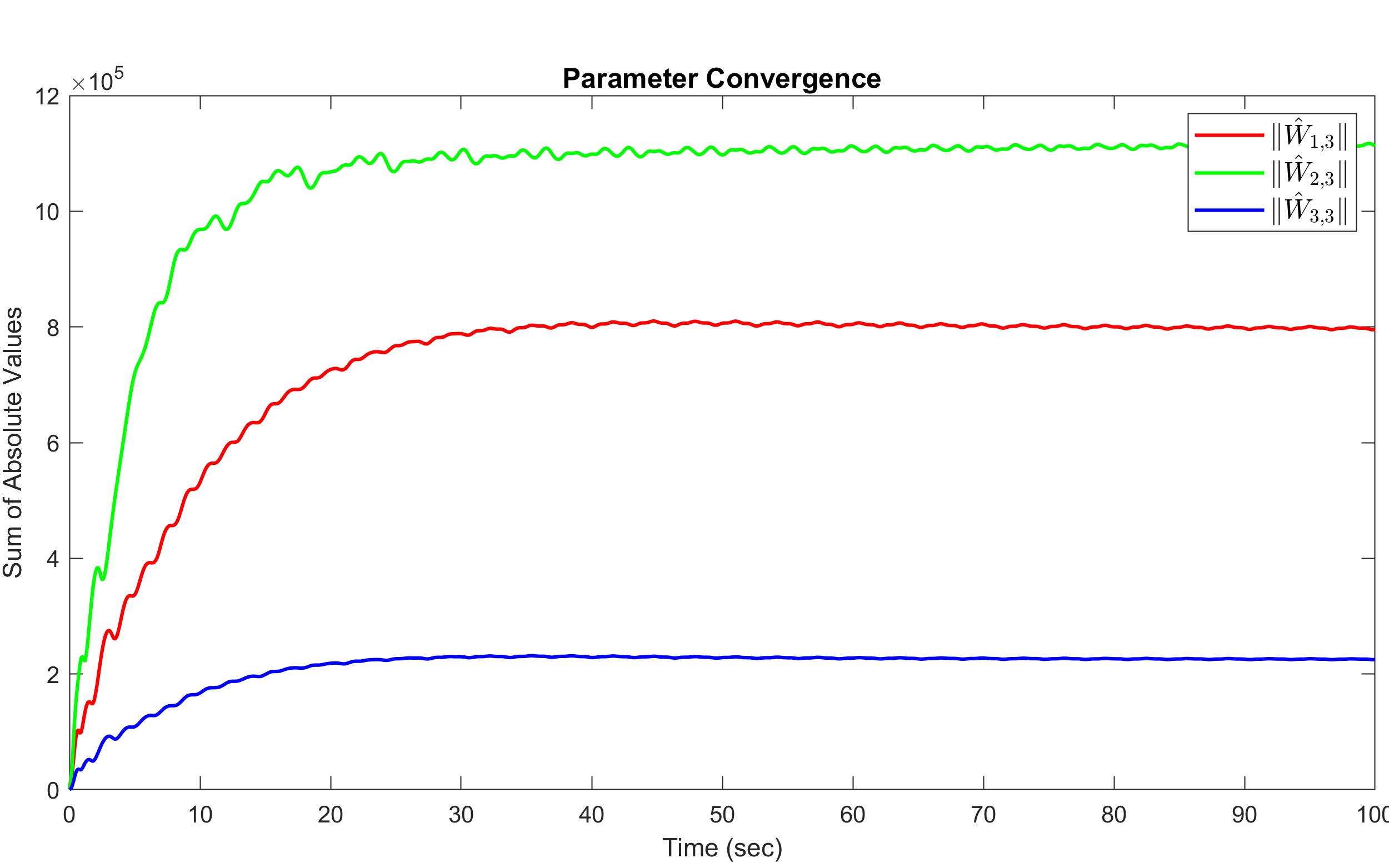}
\end{center}
\caption{$L_2$ norms of partial NN weights for AUV 3}
\label{fig:paramconv}
\end{figure}

 
The sum of the absolute values of the neural network (NN) weights in Fig.~\ref{fig:paramconv}, along with the NN weights depicted in Fig.~\ref{fig:ppconv}, demonstrates that the function approximation from the second layer is achieved accurately. This is evidenced by the convergence of all neural network weights to their optimal values during the learning process, which is in alignment with Theorem~\ref{theorem4}.
 
Fig.~\ref{fig:funcapp3} presents the neural network approximation results for the unknown system dynamics \(F_3(Z_3)\) (as defined in \eqref{eq:nonlinearfunc} for the third AUV, using a Radial Basis Function (RBF) Neural Network. The approximations are plotted for both \(W^T_{k,3} S_{k,3}(Z_3)\) and \(\bar{W}^T_{k,3} S_{k,3}(Z_3)\) for all \(k \in I[1,3]\). The results confirm that locally accurate approximations of the AUV's nonlinear dynamics are successfully achieved. Moreover, this learned knowledge about the dynamics is effectively stored and represented using localized constant RBF NNs.

\setcounter{figure}{7}
\setcounter{subfigure}{0}
\begin{subfigure}
\setcounter{figure}{7}
\setcounter{subfigure}{0}
    \centering
    \begin{minipage}[b]{0.5\textwidth}
        \includegraphics[width=\linewidth]{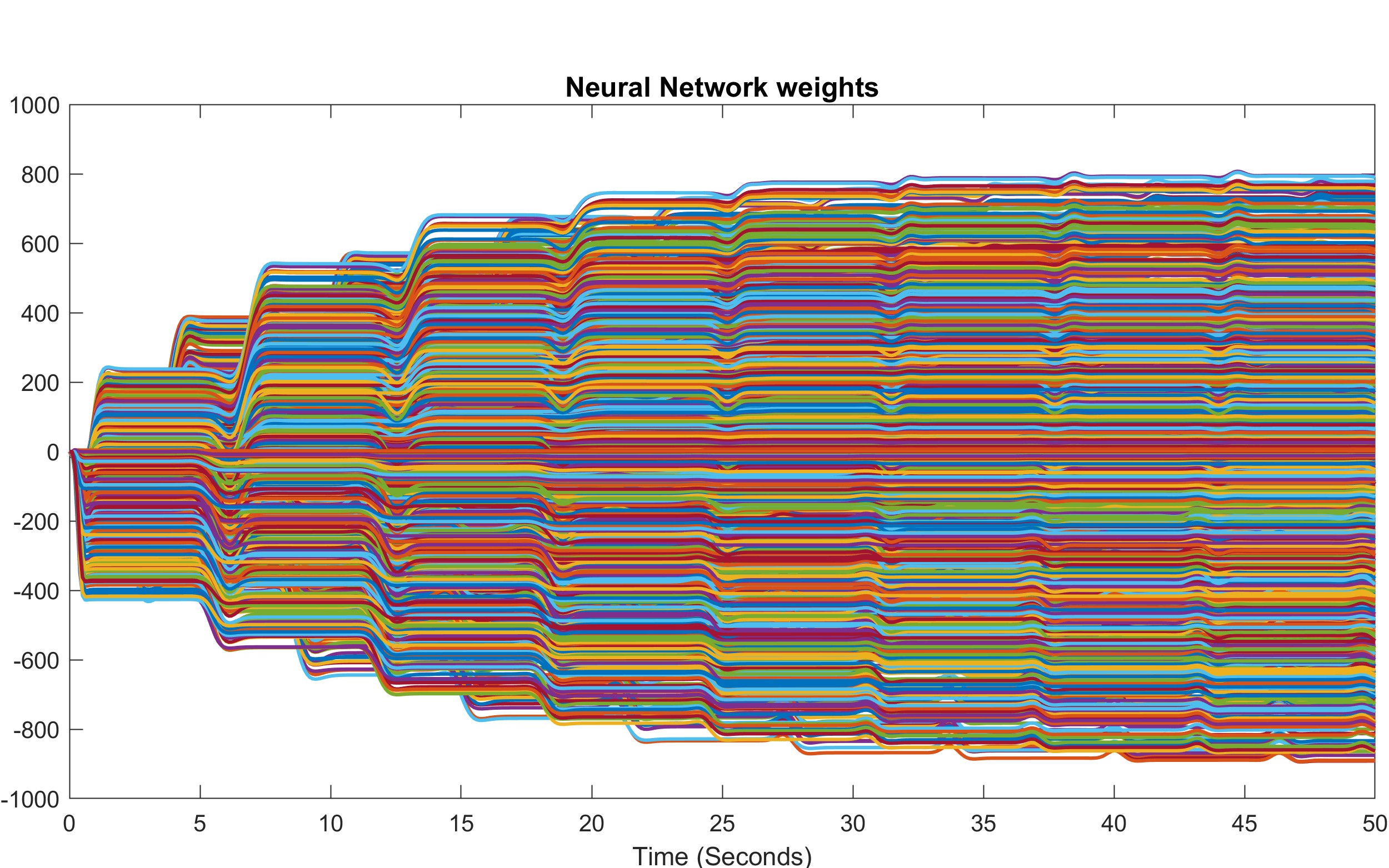}
        \caption{\(\hat{W}_{1,3}\).}
        \label{fig:ppconv1}
    \end{minipage}  
   
\setcounter{figure}{7}
\setcounter{subfigure}{1}
    \begin{minipage}[b]{0.5\textwidth}
        \includegraphics[width=\linewidth]{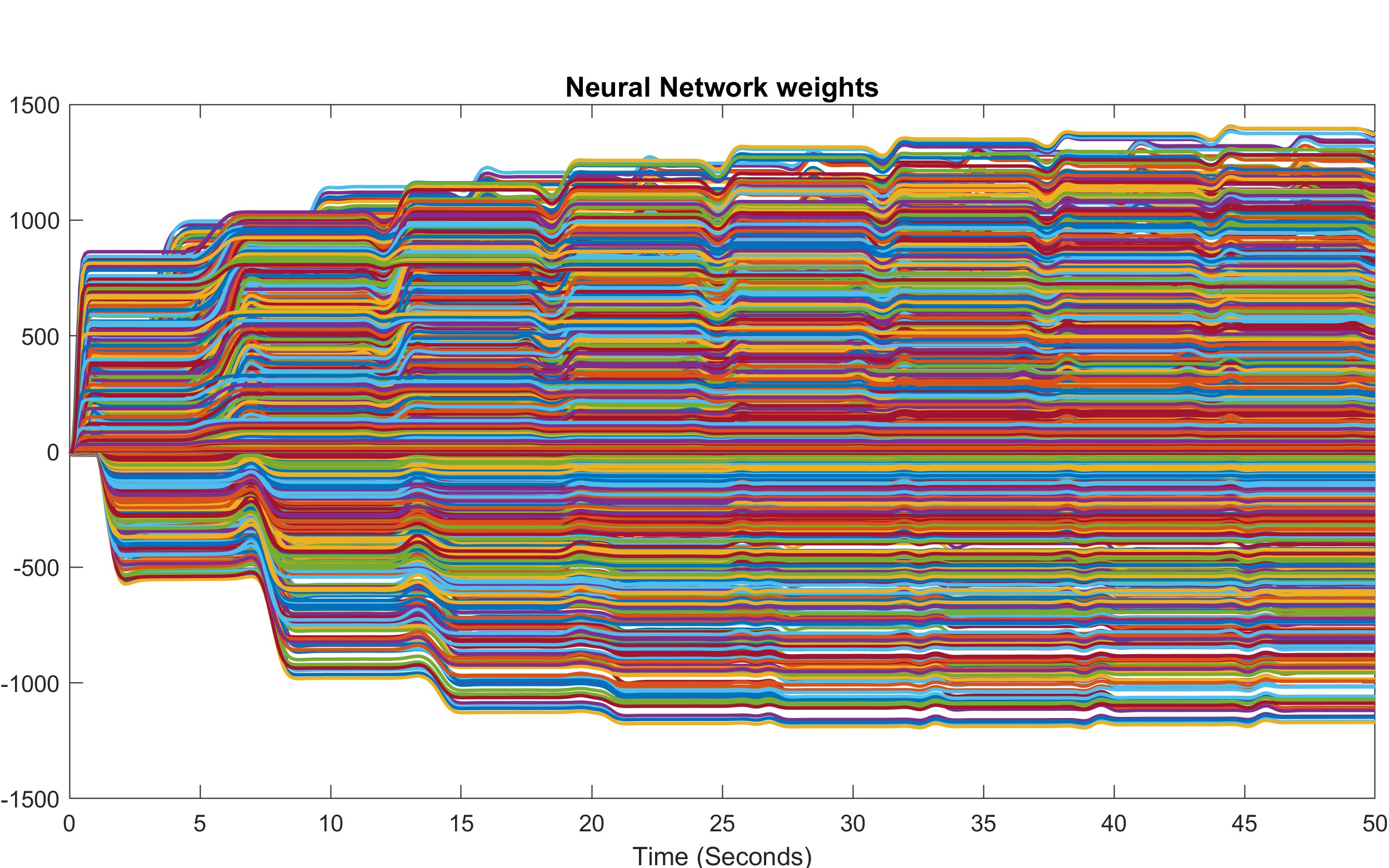}
        \caption{\(\hat{W}_{2,3}\).}
        \label{fig:ppconv2}
    \end{minipage}

\setcounter{figure}{7}
\setcounter{subfigure}{2}
    \begin{minipage}[b]{0.5\textwidth}
        \includegraphics[width=\linewidth]{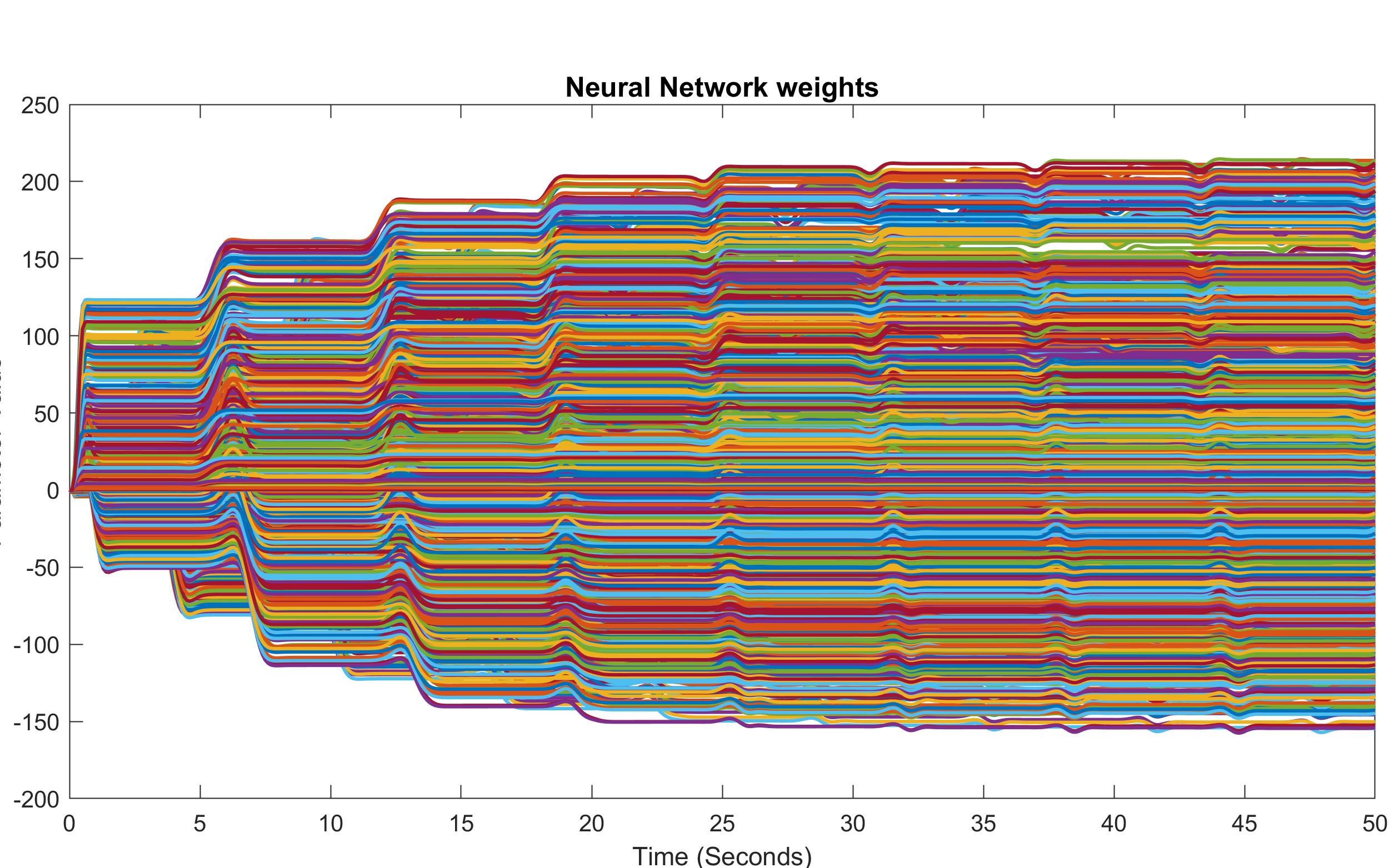}
        \caption{\(\hat{W}_{3,3}\).}
        \label{fig:ppconv3}
    \end{minipage}
    
\setcounter{figure}{7}
\setcounter{subfigure}{-1}
    \caption{Neural Network weights convergence for the third AUV (\( i = 3\)).}
    \label{fig:ppconv}
\end{subfigure}

\setcounter{figure}{8}
\setcounter{subfigure}{0}
\begin{subfigure}
\setcounter{figure}{8}
\setcounter{subfigure}{0}
    \centering
    \begin{minipage}[b]{0.5\textwidth}
        \includegraphics[width=\linewidth]{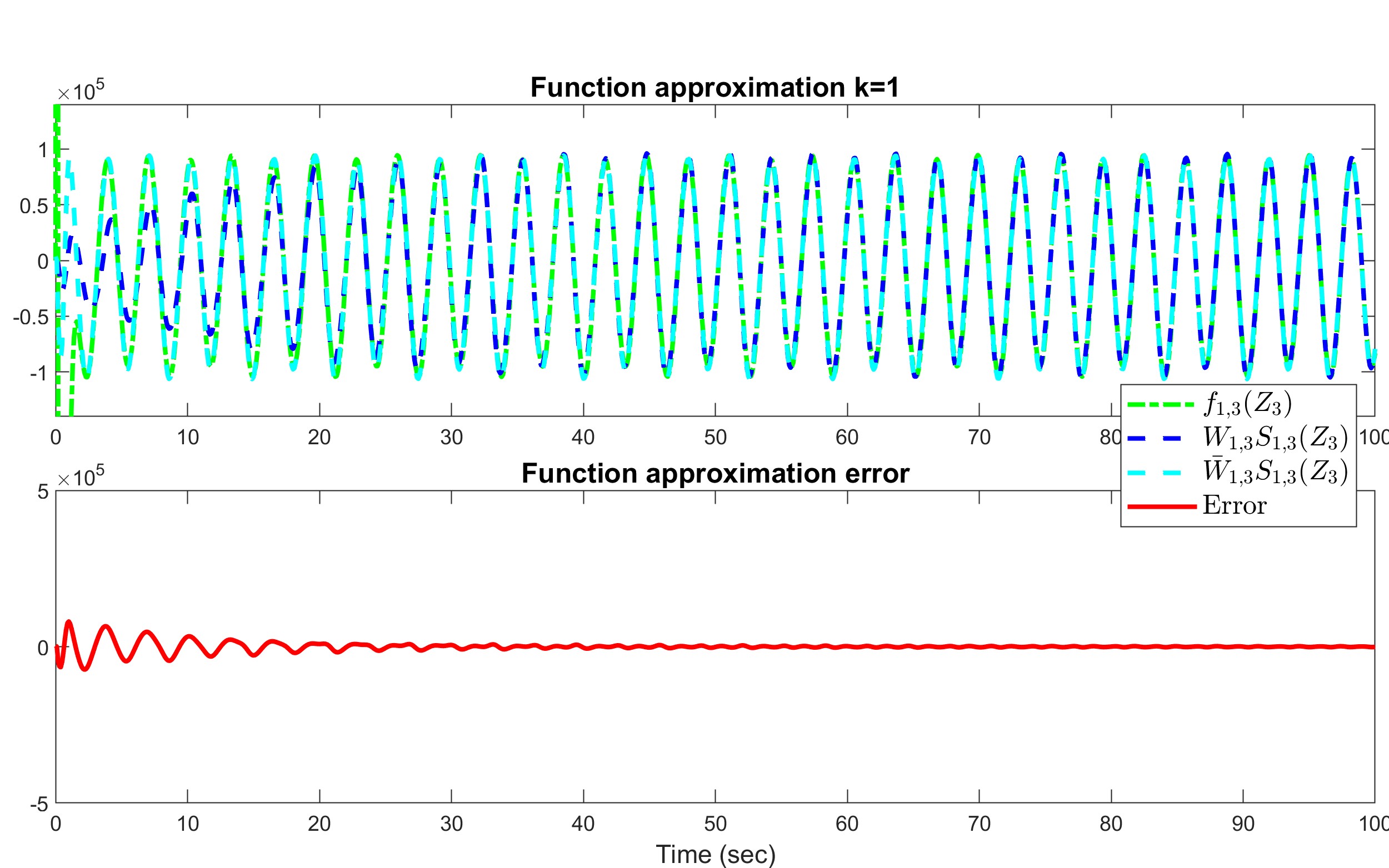}
        \caption{\(k=1\).}
        \label{fig:funcapp3_1}
    \end{minipage}  
   
\setcounter{figure}{8}
\setcounter{subfigure}{1}
    \begin{minipage}[b]{0.5\textwidth}
        \includegraphics[width=\linewidth]{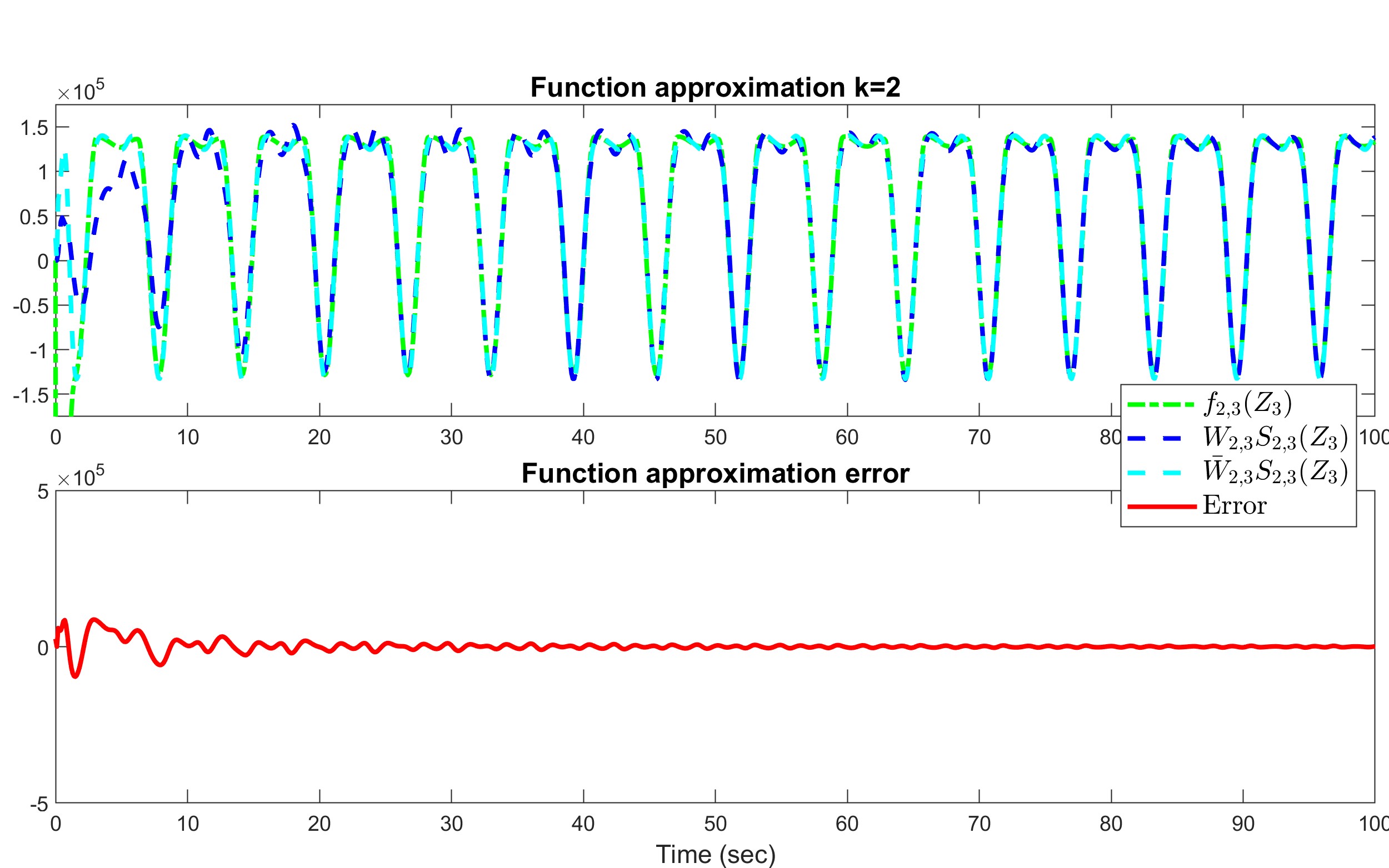}
        \caption{\(k=2\).}
        \label{fig:funcapp3_2}
    \end{minipage}

\setcounter{figure}{8}
\setcounter{subfigure}{2}
    \begin{minipage}[b]{0.5\textwidth}
        \includegraphics[width=\linewidth]{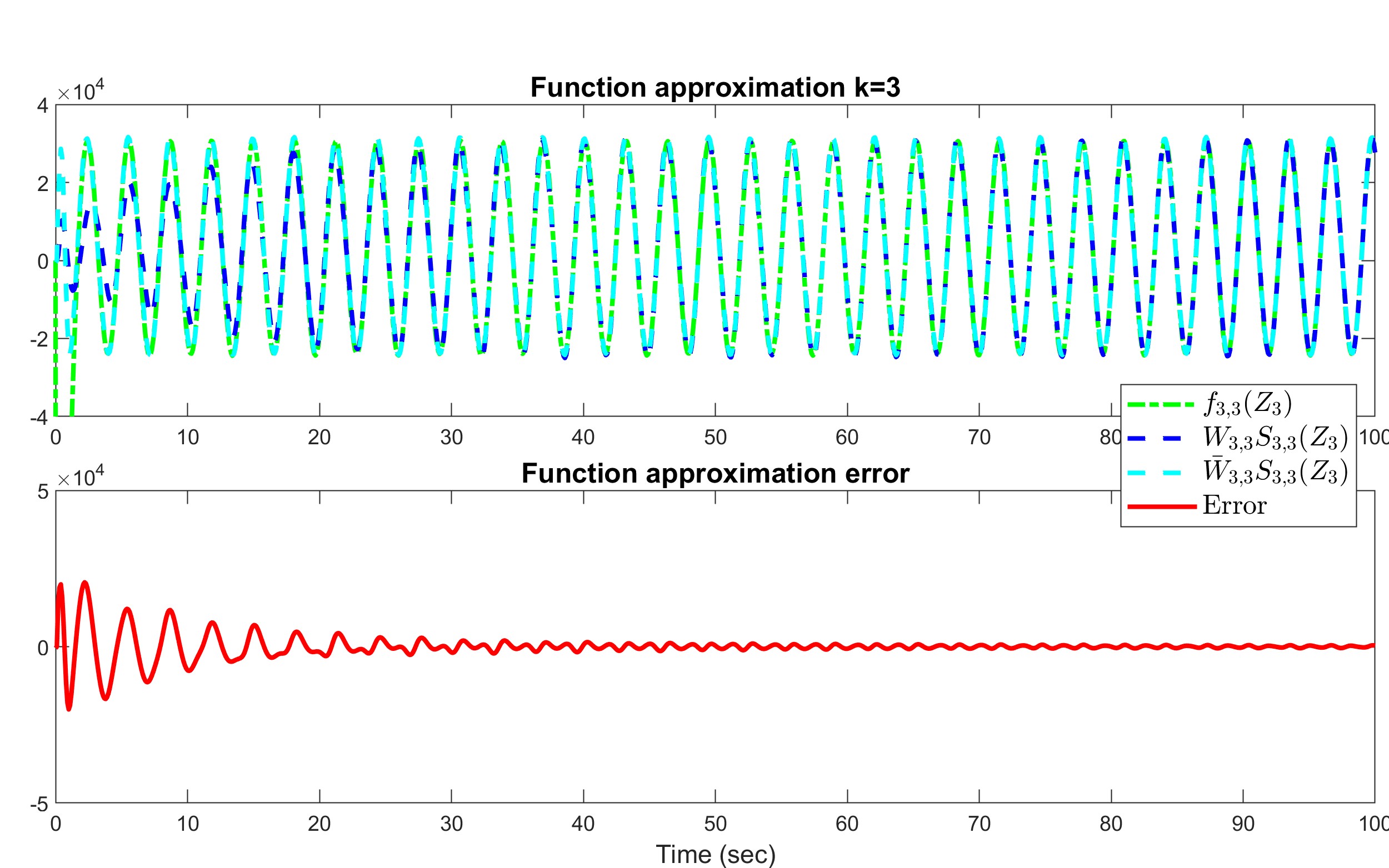}
        \caption{\(k=3\).}
        \label{fig:funcapp3_3}
    \end{minipage}
    
\setcounter{figure}{8}
\setcounter{subfigure}{-1}
    \caption{Function approximation for \(k = 1, 2, 3\) for the \(3^{rd}\) AUV (\(i = 3\)): Comparison of \(\bar{W}_{k,3}^{T}S_{k,3}(Z_{3})\), \({W}_{k,3}^{T}S_{k,3}(Z_{3})\), and \(f_{k,3}(Z_{3})\).}
    \label{fig:funcapp3}
\end{subfigure}

\subsection{Simulation for Formation Control with Pre-learned Dynamics}
\label{sec:formationsimulation}
To evaluate the distributed control performance of the multi-AUV system, we implemented the pre-learned distributed formation control law. This strategy integrates the estimator observer \eqref{eqn:DAO} and \eqref{eq:adaptationlaw}, this time coupled with the constant RBF NN controller \eqref{42}. We employed the virtual leader dynamics described in \eqref{eqn:leaderdynamicwithinitials} to generate consistent position tracking reference signals, as previously discussed in Section~\ref{sec:DDLsimulation}. To ensure a fair comparison, identical initial conditions and control gains and inputs were used across all simulations. Fig~\ref{fig:trackconte} illustrates the comparison of the tracking control results from \eqref{eq:ddlfeedback} and \eqref{eq:robustlaw} with the results using pre-trained weights \(\bar{w}\) in \eqref{42}.

\setcounter{figure}{9}
\setcounter{subfigure}{0}
\begin{subfigure}
\setcounter{figure}{9}
\setcounter{subfigure}{0}
    \centering
    \begin{minipage}[b]{0.5\textwidth}
        \includegraphics[width=\linewidth]{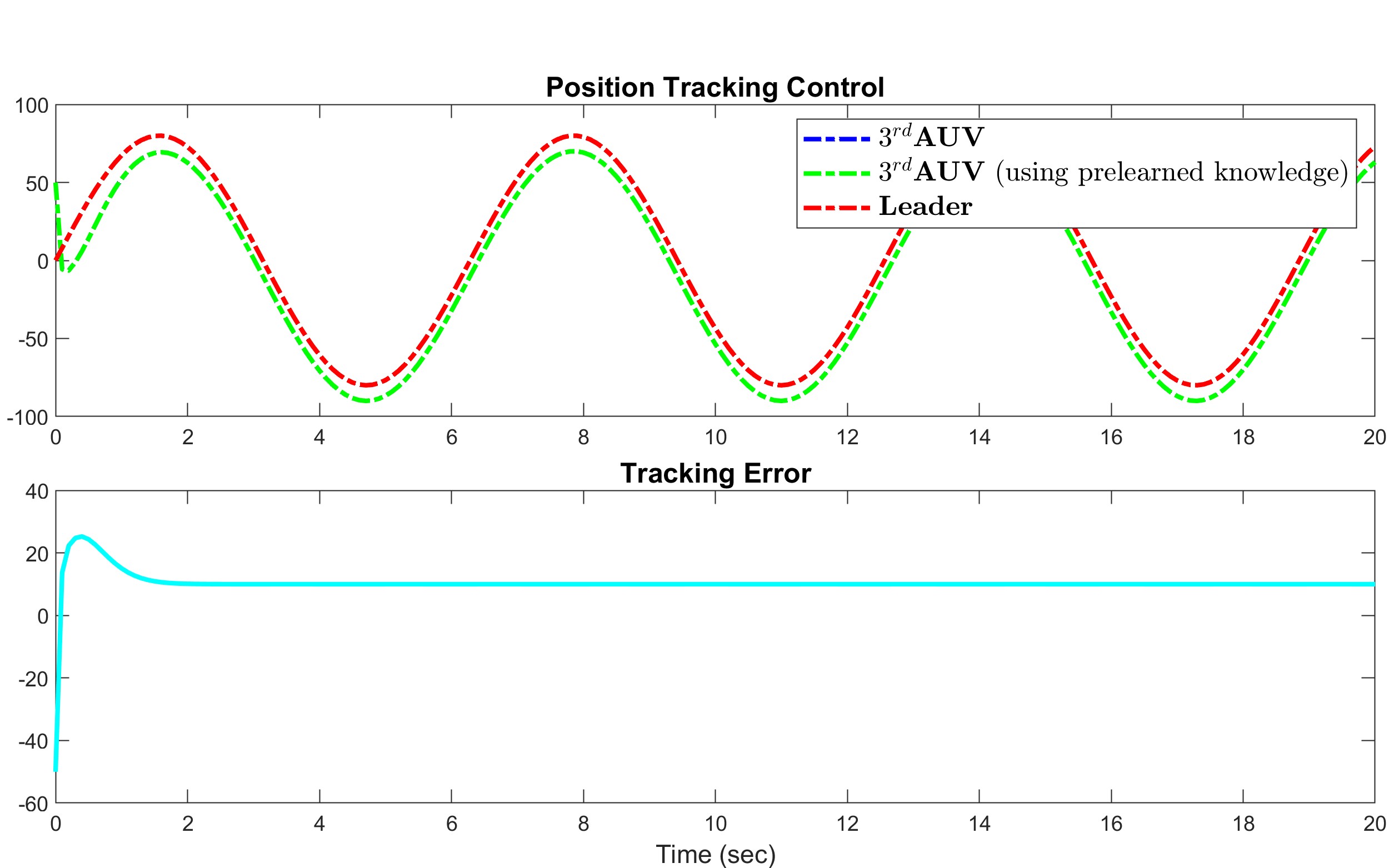}
        \caption{\(x_i \rightarrow x_0 \, \text{(m)}\).}
        \label{fig:trackconte1}
    \end{minipage}  
   
\setcounter{figure}{9}
\setcounter{subfigure}{1}
    \begin{minipage}[b]{0.5\textwidth}
        \includegraphics[width=\linewidth]{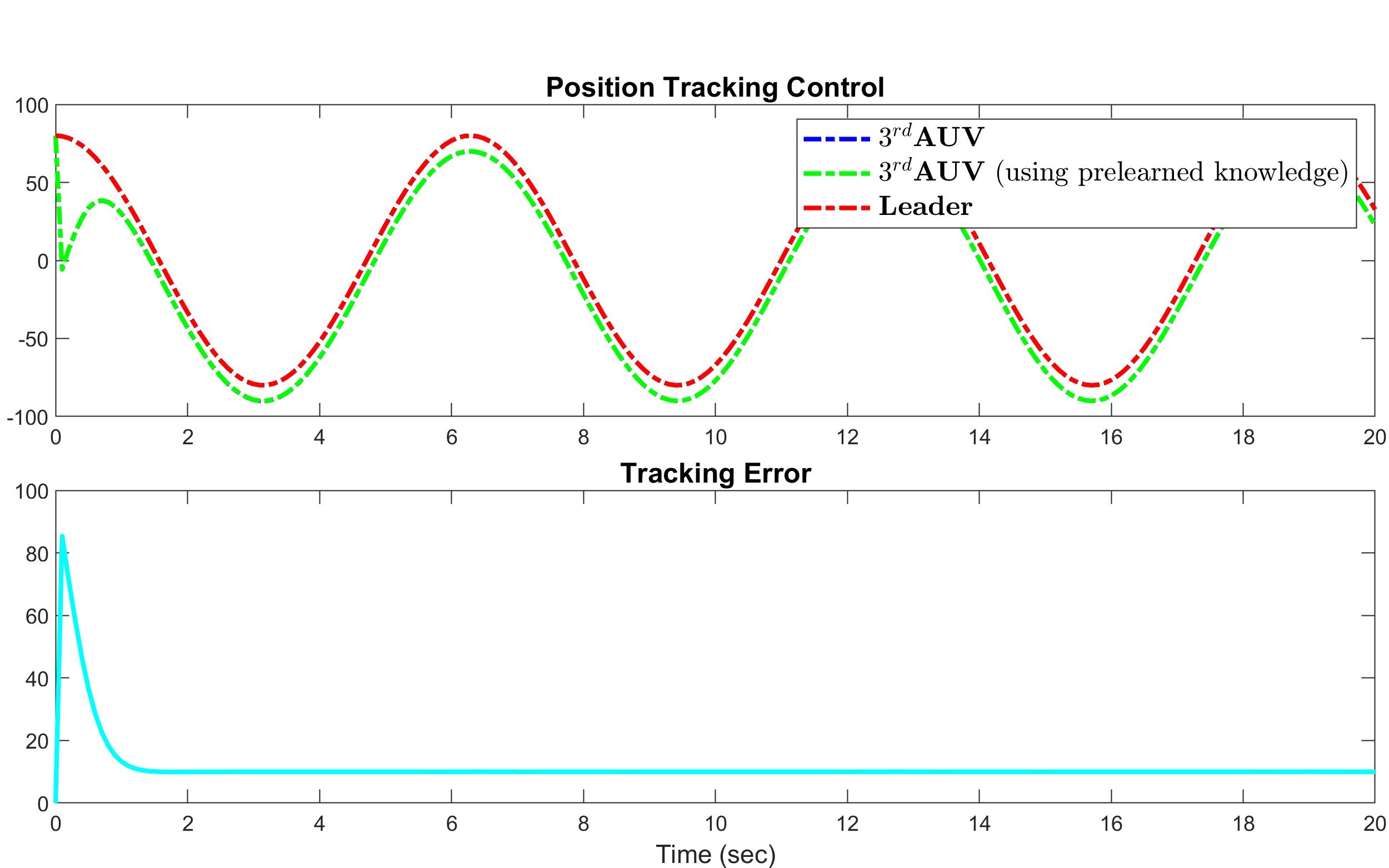}
        \caption{\(y_i \rightarrow y_0 \, \text{(m)}\).}
        \label{fig:trackconte2}
    \end{minipage}

\setcounter{figure}{9}
\setcounter{subfigure}{2}
    \begin{minipage}[b]{0.5\textwidth}
        \includegraphics[width=\linewidth]{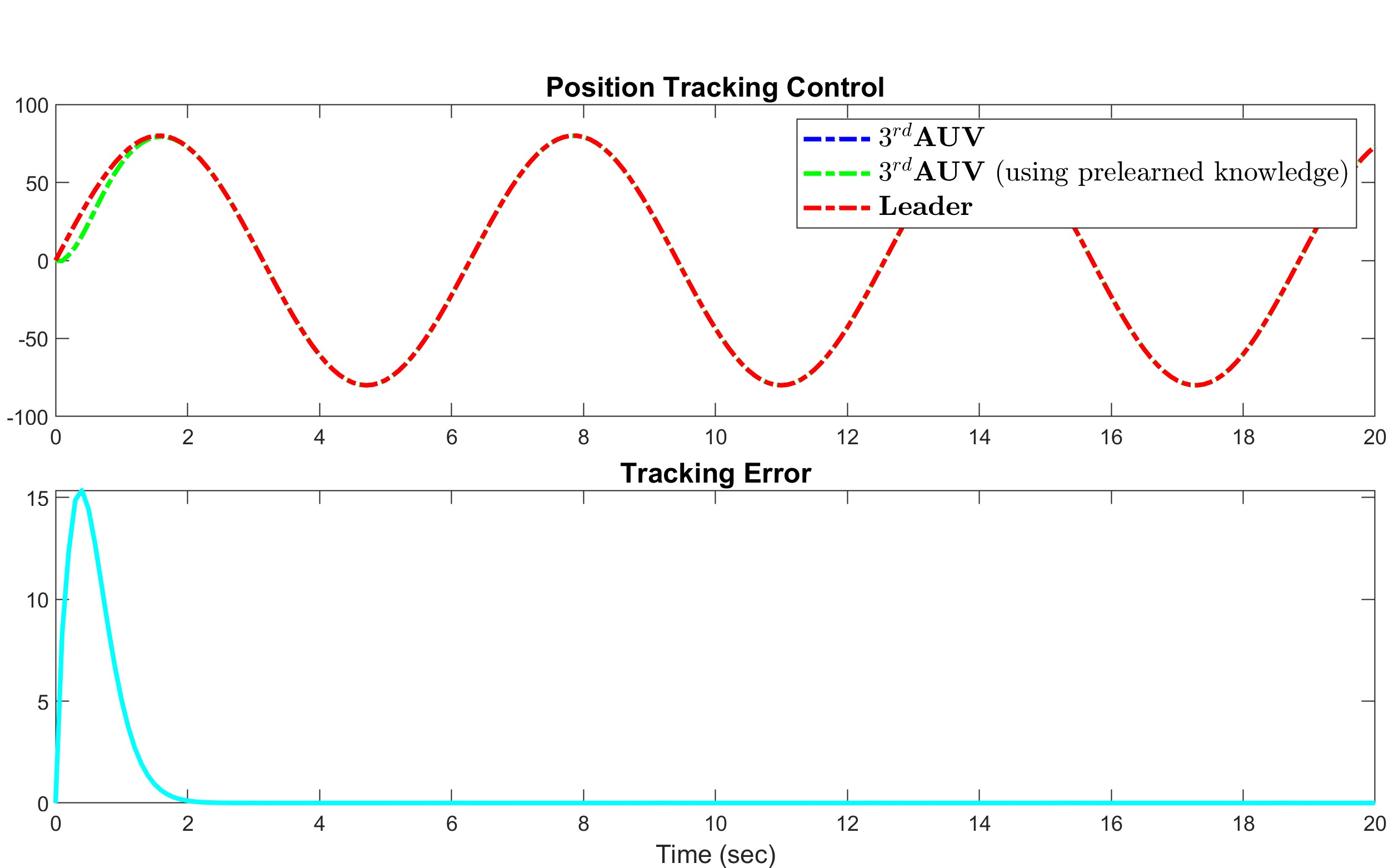}
        \caption{\(\psi_i \rightarrow \psi_0 \, \text{(deg)}\).}
        \label{fig:trackconte3}
    \end{minipage}
    
\setcounter{figure}{9}
\setcounter{subfigure}{-1}
    \caption{Postition Tracking Control using pretrained weights ($\bar{w}$)}
    \label{fig:trackconte}
\end{subfigure}

The control experiments and simulation results presented demonstrate that the constant RBF NN control law \eqref{eq:adaptationlaw} can achieve satisfactory tracking control performance comparable to that of the adaptive control laws \eqref{eq:ddlfeedback} and \eqref{eq:robustlaw}, but with notably less computational demand. The elimination of online recalculations or readaptations of the NN weights under this control strategy significantly reduces the computational load. This reduction is particularly advantageous in scenarios involving extensive neural networks with a large number of neurons, thereby conserving system energy and enhancing operational efficiency in real-time applications.

For the design of this framework, all system dynamics are assumed to be unknown, which allows its application to any AUV, regardless of environmental conditions. This universality includes adaptation to variations such as water flow, which can increase the AUV's effective mass through the added mass phenomenon, as well as influence the AUV's inertia. Additionally, buoyancy force, which varies with depth, impacts the inertia. The lift force, dependent on varying water flows or currents and the AUV's shape or appendages, along with the drag force from water viscosity, significantly influences the damping matrix in the AUV's dynamics. These factors ensure the framework's robustness across diverse operational scenarios, allowing effective operation across a range of underwater conditions including those affected by hydrodynamic forces and torques.

The controller's design enables it to maintain stability and performance in dynamic and often unpredictable underwater environments. The stored weights from the training process enable the AUVs to recall and apply learned dynamics efficiently, even after the systems are shut down and restarted. This feature ensures that AUVs equipped with our control system can quickly resume operations with optimal control settings, regardless of environmental changes or previous operational states. Our approach not only demonstrates a significant advancement in the field of autonomous underwater vehicle control but also establishes a foundation for future enhancements that could further minimize energy consumption and maximize the adaptability and resilience of AUV systems in challenging marine environments

\section{Conclusion}
\label{sec:conclusion}
In conclusion, this paper has introduced a novel two-layer control framework designed for Autonomous Underwater Vehicles (AUVs), aimed at universal applicability across various AUV configurations and environmental conditions. This framework assumes all system dynamics to be unknown, thereby enabling the controller to operate independently of specific dynamic parameters and effectively handle any environmental challenges, including hydrodynamic forces and torques. The framework consists of a first-layer distributed observer estimator that captures the leader's dynamics using information from adjacent agents, and a second-layer decentralized deterministic learning controller. Each AUV utilizes the estimated signals from the first layer to determine the desired trajectory, simultaneously training its own dynamics using Radial Basis Function Neural Networks (RBF NNs). This innovative approach not only sustains stability and performance in dynamic and unpredictable environments but also allows AUVs to efficiently utilize previously learned dynamics after system restarts, facilitating rapid resumption of optimal operations. The robustness and versatility of this framework have been rigorously confirmed through comprehensive simulations, demonstrating its potential to significantly enhance the adaptability and resilience of AUV systems. By embracing total uncertainty in system dynamics, this framework establishes a new benchmark in autonomous underwater vehicle control and lays a solid groundwork for future developments aimed at minimizing energy use and maximizing system flexibility.
The authors declare that the research was conducted in the absence of any commercial or financial relationships that could be construed as a potential conflict of interest.
\section*{Author Contributions}
Emadodin Jandaghi and Chengzhi Yuan led the research and conceptualized the study. Emadodin Jandaghi developed the methodology, conducted the derivation, design, implementation of the method, and simulations, and wrote the manuscript. Chengzhi Yuan, Mingxi Zhou, and Paolo Stegagno contributed to data analysis and provided insights on the control and oceanographic aspects of the research. Chengzhi Yuan supervised the project, assisted with securing resources, and reviewed the manuscript. All authors reviewed and approved the final version of the manuscript.
\section*{Funding}
This work is supported in part by the National Science Foundation under Grant CMMI-1952862 and CMMI-2154901.

\bibliographystyle{Frontiers-Harvard} 
\bibliography{root}

\end{document}